\documentclass[aps,prd,twocolumn]{revtex4}

\usepackage[T1]{fontenc}
\usepackage{bm,exscale,latexsym, amsmath, amssymb, mathrsfs, dsfont,accents,tensor} 
\usepackage[dvips]{graphicx}

\newcommand{\La}{\ensuremath{\mathcal L}}

\newcommand{\A}{\ensuremath{\mathcal A}}
\newcommand{\F}{\ensuremath{\mathcal F}}
\newcommand{\M}{\ensuremath{\mathcal M}}
\newcommand{\cov}{\ensuremath{\text D}}
\newcommand{\covgr}{\ensuremath{\bar{\cal D}}}
\newcommand{\Kr}{\ensuremath{\mathscr R}}

\newlength{\bredde}
\def\slash#1{\settowidth{\bredde}{$#1$}\ifmmode\,\raisebox{.15ex}{/}
\hspace*{-\bredde} #1\else$\,\raisebox{.15ex}{/}\hspace*{-\bredde} #1$\fi}

\begin{document}
\begin{minipage}{2\linewidth}
\begin{flushright}
SI-HEP-2007-13 \\[0.2cm]
revised: April 10, 2008
\end{flushright}
\end{minipage}

\vspace{0.25in}
\title{Effective Field Theory of Gravity:\\Leading Quantum Gravitational Corrections to Newtons and Coulombs Law}
\author{\sc{Sven Faller}}
\email[]{faller@hep.physik.uni-siegen.de} 
\affiliation{Theoretische Physik 1, Fachbereich Physik, Universit\"at Siegen\\ D-57068 Siegen, Germany}

\begin{abstract}
In this paper we consider general relativity and its combination with
scalar quantum electrodynamics (QED) as an effective quantum field
theory at energies well below the Planck scale. This enables us to 
compute the one-loop quantum corrections to the Newton and Coulomb 
potential induced by the combination of graviton and photon fluctuations. 
We derive the relevant Feynman rules and compute the nonanalytical contributions
to the one-loop scattering matrix for  charged scalars  in the
nonrelativistic limit.
In particular, we derive the post-Newtonian corrections of order $Gm/\text c^2 r$ from general relativity 
and the genuine quantum corrections of order $G\hbar/\text c^3 r^2$. 
\end{abstract}
\maketitle
%
%
\section{Introduction}
At the classical level the interaction between two charged massive particles is described by Newton's and Coulomb's law,
in terms of the nonrelativistic potential
\begin{equation}\label{eq:1}
 V(r) = - G \ \frac{m_1 m_2}{r} + \frac{1}{4\pi} \frac{e_1 e_2}{r}  \; .
\end{equation}
This potential will be modified by relativistic and quantum corrections. Radiative corrections to the
Coulomb potential can be systematically calculated within QED.  However, it is well known that 
general relativity, including scalar
\cite{Hoo74, Vel76, Gor86}, fermion or photon fields \cite{Des74a, Des74b, Des74c}
is not renormalizable.
Still, on smooth-enough background space-time the gravitational field 
can be quantized consistently 
\cite{Hoo74, Vel76}, and as suggested by Donoghue \cite{Don94a, Don94b}, 
one can consider general
relativity as an effective quantum field theory  in the low-energy limit 
where renormalizability is no longer an issue.

Within this framework, the one-loop corrections to quantum gravity 
coupled to a scalar Klein-Gordon field
have already been discussed by independent groups, e.g.
Donoghue \cite{Don94a, Don94b, Don95}, Hamber and Liu \cite{Ham95}, 
Akhundov et al.~\cite{Akh97, Akh06}, and Bjerrum-Bohr et al.~\cite{Bje02, Bje03, Bje03a},
with differing results. The generalization to scalar QED has been first
discussed by Bjerrum-Bohr \cite{Bje02}. Butt \cite{But06} has also treated the case of charged fermions. 

In this paper, we are going to reconsider the results for the case
 of scalar QED coupled to gravity (SQED), in order to
perform and independent check of the above results and
 to resolve the discrepancies between \cite{Don94b, Don94a, Don95}, 
\cite{Ham95}, \cite{Akh97, Akh06} and \cite{Bje02, Bje03, Bje03a}.
 To this end, we discuss in  Sec.~\ref{Q-GR-SQED}
how to quantize general relativity/SQED with the background field method.
In the next section we review how  ultraviolet divergences from one-loop diagrams can be
absorbed into higher-dimensional operators in the effective theory.
The definition of the scattering matrix and of  the potential for  (charged or neutral) scalar particles
 is given in Sec.~\ref{S-M-PD}.
In Sec.~\ref{R-F} we perform the calculations of the relevant Feynman diagrams and
construct the scattering potentials at one loop. We compare our results
with \cite{Don94b, Don94a, Akh97, Akh06, Bje03} and \cite{Bje02}, before
we conclude with a short summary.
The Feynman rules, one-loop integrals, and details of the Fourier transformations
are presented in the Appendices \ref{A-A}, \ref{A-B} and \ref{A-C}.
%
%
\section{Quantization of General Relativity and Scalar QED}\label{Q-GR-SQED}
The Einstein-Hilbert action for the vacuum is given by
\begin{equation}\label{eq:2}
 \mathscr  S_\text{vac}= \int d^4 x \ \sqrt{-g} \ \frac{2}{\kappa^2} \ \Kr \; ,
\end{equation}
where $\kappa^2= 32 \pi G$ is the gravitational coupling,
$g_{\mu\nu}$ is the metric tensor with its determinant
$g = \textsf{det} g_{\mu\nu}$,  and
$\Kr = g^{\mu\nu} \mathcal R_{\mu\nu}$ is the curvature scalar.
In the following we use the same convention as Refs.~\cite{Don94a, Don94b, Don95, Akh97, Bje02, Bje03, Bje03a, But06, Akh06},
with the Minkowski metric  $\eta_{\mu\nu}= \textsf{diag}(+1, -1,-1,-1)$  and  $\hbar = \text c = 1$.

Additional fields can be included by adding a covariant term
 $\sqrt{-g} \ \mathcal L_m$ to the action Eq.~\eqref{eq:2},
\begin{equation}\label{eq:3}
 \mathscr S = \int d^4x \ \sqrt{-g} \ \biggl \lbrack \frac{2 \ \mathscr R}{\kappa^2} + \mathcal L_m \biggr \rbrack \; . 
\end{equation}
For instance, massive scalar fields are described by the  scalar Lagrangian
\begin{equation}\label{eq:4}
 \La_m = \frac{1}{2} \bigl( g^{\mu\nu}\  \partial_\mu \phi \ \partial_\nu \phi^\ast - m^2\ | \phi |^2 \bigr) \ .
\end{equation}
%

\subsection{Background Field Method}

Following Veltman and 't Hooft~\cite{Hoo74, Vel76} we can expand the gravitational field $g_{\mu\nu}$
into the classical background field $\bar g_{\mu\nu}$ and a small quantum fluctuation $\kappa \, h_{\mu\nu}$:
\begin{equation}\label{eq:5}
\begin{split}
 g_{\mu\nu} &= \bar g_{\mu\nu} + \kappa \, h_{\mu\nu} \ ,\\
g^{\mu\nu}&= \bar g^{\mu\nu} - \kappa \, h^{\mu\nu} + \kappa^2 \, h^\mu_\lambda \, h^{\lambda\nu} \ .
\end{split}
\end{equation}
%
Expanding the curvature to second order in the quantum field $h$, using 
\begin{equation}\label{eq:6}
 \sqrt{-g} = \sqrt{-\bar g} \biggl \lbrace 1 + \frac{\kappa}{2} h^\alpha_\alpha - \frac{\kappa^2}{4}h^\alpha_\beta h^\beta_\alpha + \frac{\kappa^2}{8}\bigl( h^\alpha_\alpha\bigr)^2 + \dots \biggr \rbrace \ ,
\end{equation}
we obtain for the gravitational part of the  Lagrangian
\begin{equation}\label{eq:7}
 \La_\text{grav}= \sqrt{- \bar g} \biggl \lbrack \frac{2}{\kappa^2} \bar{\Kr} + \La_{\text{grav}}^{(1)} + \La_{\text{grav}}^{(2)} + \dots \biggr \rbrack  
\end{equation}
with
\begin{equation}\label{eq:8}
\begin{split}
\La_{\text{grav}}^{(1)} &= \frac{1}{\kappa} \, h_{\mu\nu}  \, \bigl \lbrack \bar g^{\mu\nu} \bar{\Kr} - 2 \bar{\Kr}^{\mu\nu} \bigr \rbrack \; ,\\
\La_{\text{grav}}^{(2)} &= \frac{1}{2} \covgr_\alpha h_{\mu\nu} \, \covgr^\alpha h^{\mu\nu} - \frac{1}{2} \covgr_\alpha h \, \covgr^\alpha h\\   	&+ \covgr_\alpha h \, \covgr_\beta h^{\alpha \beta}- \covgr_\alpha h_{\mu\beta} \, \covgr^\beta h^{\mu\alpha} \\
	&+ \bar{\Kr} \biggl( \frac{1}{4} h^2 - \frac{1}{2} h_{\mu\nu} h^{\mu\nu} \biggr) \\
	&+ \bar{\Kr}^{\mu\nu} \bigl( 2 h^\lambda_\mu h_{\nu\lambda} - h \, h_{\mu\nu} \bigr) \; .
\end{split}
\end{equation}
Here the curvature scalar $\bar \Kr$ and the covariant derivative $\covgr_\alpha$ 
are evaluated from the background metric $\bar g_{\mu\nu}$.

In order to quantize gravity we have to fix the gauge and introduce a Faddeev-Popov ghost field.
From \cite{Hoo74} we take the gauge fixing Lagrangian 
\begin{equation}\label{eq:9}
 \La_\text{gf}= \sqrt{-\bar g} \biggl \lbrace \biggl( \covgr^\nu h_{\mu\nu} - \frac{1}{2} \covgr_\mu h \biggr)\biggl(\covgr_\lambda h^{\mu\lambda} - \frac{1}{2} \covgr^\mu h\biggr) \biggr \rbrace \ , 
\end{equation}
and  the ghost Lagrangian
\begin{equation}\label{eq:10}
 \La_\text{ghost} = \sqrt{-\bar g} \, \eta^{\ast\mu} \bigl \lbrack \covgr_\lambda \covgr^\lambda \, g_{\mu\nu} - \bar\Kr_{\mu\nu} \bigr \rbrack \, \eta^\nu \ ,
\end{equation}
with $\eta^\nu$ being the complex Faddeev-Popov ghost field. 

From this Lagrangian one derives the graviton propagator as quoted in Appendix~\ref{A-A}.
%
%
\subsection{Combining Scalar QED and General Relativity}
 
The generally covariant Lagrangian for scalar QED is \cite{Bje02}
\begin{equation}\label{eq:11}
 \begin{split}
  \La_\text{SQED} = \sqrt{-g} \biggl \lbrack &-\frac{1}{4} \bigl( g^{\alpha\mu}g^{\beta\nu} \F_{\alpha\nu} \F_{\mu\beta} \bigr) \\
	&+ g^{\mu\nu} \, \cov_\mu \phi \ \cov_\nu \phi^\ast - m^2 |\phi|^2 \biggr \rbrack \; ,
 \end{split}
\end{equation}
with the photon field strength $\F_{\mu\nu} = \cov_\mu \A_\nu - \cov_\nu \A_\mu$, 
and the QED covariant derivative $\cov_\mu = \partial_\mu - i e_q \A_\mu (x)$,
and $e_q$ being the charge of the scalar particle. 
In order to get the relevant Feynman rules at one-loop order, we have to expand the Lagrangian
$\La_\text{SQED}$ to first order in the quantum field $h$. Using Eqs.~\eqref{eq:5},
\eqref{eq:7} and working with the Lorenz gauge for the photon field,
\begin{equation}\label{eq:12}
 \La^{\text C} = - \frac{1}{2} \ (\partial_\mu \A^\mu )^2 \; ,
\end{equation}
we get for the photon part
\begin{equation}\label{eq:13}
 \begin{split}
  \La_\text{photon}^{(1)} =& \sqrt{- \bar g} \biggl \lbrace - \frac{\kappa h}{4} \bigl( \partial_\mu  \A_\nu \partial^\mu \A^\nu - \partial_\mu \A_\nu \partial^\nu \A^\mu \bigr) \\
	&+ \frac{\kappa}{2} \ h^{\mu\nu} \ \bigl( \partial_\mu \A_\alpha \partial_\nu \A^\alpha + \partial_\alpha \A_\mu \partial^\alpha \A_\nu \\
	&- \partial_\alpha \A_\mu \partial_\nu \A^\alpha - \partial_\alpha \A_\nu \partial_\mu \A^\alpha \bigr) \biggr \rbrace \; ,
 \end{split} 
\end{equation}
and for the matter Lagrangian
\begin{align}
    \La_\text{m}^{(1)} =& \frac{1}{2}\ \kappa h \bigl( |\partial_\mu \phi |^2 - m^2 |\phi|^2 \bigr) - \kappa h^{\mu\nu} \bigl( \partial_\mu \phi^\ast \partial_\nu \phi \bigr)  \notag \\
	&+ i e_q \kappa \bigl ( \A_\mu \partial^\mu \phi^\ast \phi - \A_\mu \phi^\ast \partial^\mu \phi \bigr) + e_q^2 \A_\mu \A^\mu |\phi|^2\notag\\
	&+ \frac{1}{2} \kappa h \bigl \lbrack \partial_\mu \phi^\ast (i e_q \A^\mu ) \phi - (i e_q A^\mu) \phi^\ast \partial_\mu \phi \rbrack \notag\\
	&- \kappa h^{\mu\nu} \bigl \lbrack \partial_\mu \phi^\ast (ie_q \A_\nu ) \phi + (i e_q \A_\mu ) \phi^\ast \partial_\nu \phi \bigr \rbrack \label{eq:14}
 \end{align}
The Feynman rules for interaction vertices between photons, complex scalars and gravitons
can be derived from \eqref{eq:13} and \eqref{eq:14}, \cite{Bje02}, and
are presented in Appendix~\ref{A-A}.
%

\section{General Relativity and Scalar QED as an Effective Field Theory}\label{GR-SQED-EF}

The treatment of general relativity as a quantum theory has been discussed in several publications,
e.g.~\cite{Ein16, DeW67, Fad67, Schw63, Wein64, Man68, Hoo74}. 
The standard formalism would start from the action \eqref{eq:2} and consider
the metric as the gravitational field which has to be quantized in the usual way.
However, such a theory is not self-contained, since at each order in perturbation theory
loop diagrams generate new terms which are not present in the original action
 \cite{Hoo74, Vel76, Gor86, Ven92}. Because of this renormalization problem,
a consistent definition of general relativity as a \emph{fundamental} quantum
theory is unknown so far \cite{Akh06}. 

As mentioned above, the problem can be solved by considering general relativity
as an effective quantum field theory at low energies \cite{Don94a, Don94b, Don95}. 
In this case, the effective Lagrangian
for pure gravity  (without a cosmological constant)  has the expansion
\begin{equation}\label{eq:15}
 \La_\text{grav} = \sqrt{-g}\biggl \lbrace \frac{2 \Kr }{\kappa^2} + a_1 \Kr^2 + a_2 \mathcal R_{\mu\nu} \mathcal R^{\mu\nu} + \dots \biggr \rbrace \; .
\end{equation}
Here, the coefficients $a_1$ and $a_2$ are energy scale dependent coupling constants in
the effective theory which -- in principle -- could be determined by an experimental measurement.
The ellipses denote higher-derivative couplings which are even more suppressed at low energies. 

Similarly, as shown in \cite{Bje02} general relativity and scalar QED can be 
described in the effective-theory framework as well.
The 1-loop singularities from graviton and
photon exchange can be absorbed into the Lagrangian
\begin{equation}\label{eq:16}
\Delta \La_\text{photon} = \sqrt{-g} \biggl \lbrace c_1 T^2_{\mu\nu} + c_2 \mathcal R_{\mu\nu} T^{\mu\nu} + \dots \biggr \rbrace \,,
\end{equation}
where $T_{\mu\nu} = \mathcal F_{\mu\alpha} \mathcal F^\alpha{}_\nu - \frac{1}{4} g_{\mu\nu} \mathcal F^{\alpha\beta} \mathcal F_{\alpha\beta}$ is the Maxwell stress tensor \cite{Des74a, Des74b, Bje02}.
For the matter Lagrangian describing the complex scalar fields, one obtains additional terms \cite{Bje02}
\begin{equation}\label{eq:17}
\begin{split}
 \La_\text{scalar} = \sqrt{-g} & \biggl \lbrace d_1 \mathcal R^{\mu\nu} \partial_\mu \phi^\ast \partial_\nu \phi + d_2 \Kr |\partial_\mu \phi|^2\\
	&+ d_3 \Kr m^2 |\phi|^2 + \dots \biggr \rbrace
\end{split}
\end{equation}
An explicit discussion of the coefficients  $a_1$, $a_2$, $c_1$, $c_2$, $d_1$, $d_2$
and $d_3$,  can be found in \cite{Don94a, Don94b, Don95, Bje03} and \cite{Bje02}.
%

\section{Scattering Matrix and  Potential}\label{S-M-PD}

According to \cite{Don94a,Don94b, Don95, Bje02,Bje03}, the scattering amplitude  
for  two massive, charged scalar particles as a function of the momentum transfer
 $q^2 = (k-k')^2=(p-p')^2$ can be expanded as
\begin{equation}\label{eq:18}
\begin{split}
 \mathcal M(q) \sim &\biggl \lbrack A + B q^2 + \dots + \bigl( \alpha_1 \kappa^2 + \alpha_2 e^2 \bigr) \ \frac{1}{q^2} \\
	&+ \beta_1 e^2 \kappa^2 \log (-q^2) + \beta_2 e^2 \kappa^2 \ \frac{1}{\sqrt{-q^2}} + \dots \biggr \rbrack \; ,
\end{split}
\end{equation}
where the coefficients $A,B,\dots$ and $\alpha_i,\beta_i$ depend on the particle masses $m_i$ and charges $e_i/e$.
The terms with $A, B, \dots$ in this equation are analytical in $q^2$ and correspond to local
interactions. They are relevant for large momentum transfer and are also related to 
the UV renormalization of the theory, but will not be of further interest in this work. 
Rather, we concentrate on the nonanalytical terms, parametrized by the coefficients  $\alpha_i$ and $\beta_i$, which will 
induce the nonlocal, long-ranged interactions described by the nonrelativistic potential.
The nonanalytical parts in the scattering-matrix also ensure unitarity
\cite{Don94b, Don94a, Don95, Bje03, Bje02}.

The nonrelativistic potential will be calculated from the scattering amplitude via the
spatial Fourier transform \cite{Don94b, Don94a, Don95, Ham95, Bje03, Bje02}
\begin{equation}\label{eq:19}
 V(r) = - \frac{1}{2 m_1}\frac{1}{2 m_2} \int \frac{d^3 q}{(2 \pi)^3} \ e^ {i \vec q \cdot \vec r} \ \M(\vec q) \; .
\end{equation}
The non-analytical terms in $\M$ correspond to an expansion in powers of  $\frac{1}{r}$ for $V(r)$.
For alternative definitions of the potential, see the discussion in, e.g., \cite{Muz95,Kaz01,Dal94}.
%

\section{Results for the Feynman Diagrams}\label{R-F}

In the following section we discuss the calculation of the various Feynman diagrams and
their contribution to the scattering amplitude. At the end of this section, we
summarize the one-loop result for the nonrelativistic potential which represents the main result
of this paper.
Part of the calculation for the Feynman diagrams has been done with the help of Mathematica 5.2\texttrademark
\footnote{Mathematica 5.2 is registered trademark of Wolfram Research, Inc.}
using the Ricci-application\textcopyright \footnote{Ricci version 1.51 copyright 1992 -- 2004 John M. Lee}
for tensor calculus. The Fourier transformations, necessary to obtain the potential in coordinate space,
are listed in Appendix \ref{A-C}.
%

\subsection{Tree Diagrams}

\begin{figure}[t!!]
\begin{minipage}{0.45\linewidth}
\includegraphics[scale=.75]{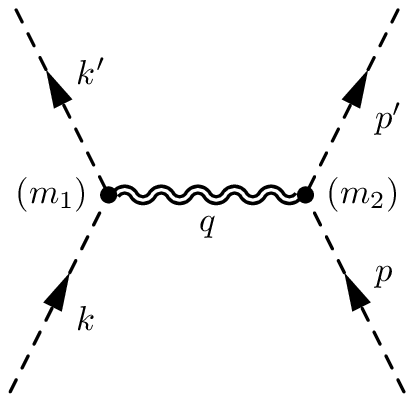}

\centering{(a)}
\end{minipage}
\begin{minipage}{0.45\linewidth}
\includegraphics[scale=.75]{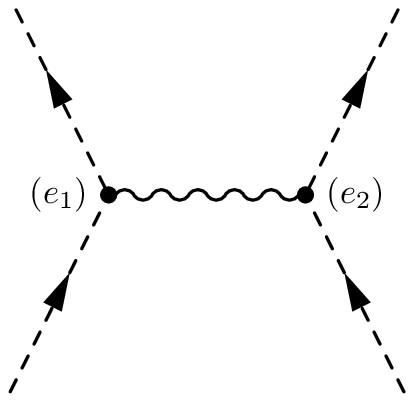}

\centering{(b)}
\end{minipage}
\caption{The two tree diagrams.}\label{tree}
\end{figure}

The tree diagrams contributing to the scattering amplitude are shown in Fig.~\ref{tree}. 
The (incoming/outgoing) momenta for the first particle are denoted as $(k/k')$ with the
(mass/charge) being $(m_1/e_1)$, and for the second particle as 
 $(p/p')$ and $(m_2/e_2)$, respectively. 
They have the form
\begin{equation}\label{eq:20}
 i \M_{1(\text a)}(q) = \tau_1^{\mu\nu} (k, k',m_1)  \biggl \lbrack \frac{i \mathcal P_{\mu\nu\alpha\beta}}{q^2}\biggr \rbrack  \tau_1^{\alpha\beta} (p,p',m_2)
\end{equation}
and
\begin{equation}\label{eq:21}
 i \M_{1 (\text b)}(q) = \tau_2^\gamma (k,k',e_1)  \biggl \lbrack \frac{-i \ \eta_{\gamma \delta}}{q^2}\biggr \rbrack  \tau_2^\delta (p,p',e_2) \; . 
\end{equation}
where the quantities $\tau_i$ and $\mathcal P_{\mu\nu\alpha\beta}$ are given in Appendix~\ref{A-A}.
After Fourier transformation we recover the Newtonian nonrelativistic potential
\begin{equation}\label{eq:22}
 V_{1(\text a)}(r) = - G \, \frac{ m_1  m_2}{r} 
\end{equation}
and the classical Coulomb potential
\begin{equation}\label{eq:23}
 V_{1(\text{b})} (r)= \frac{1}{4\pi}\frac{e_1 e_2}{ r} \; .
\end{equation}
%

\subsection{Vertex Corrections}

\begin{figure}[t!!]
\begin{minipage}{0.45\linewidth}
\includegraphics[scale=.75]{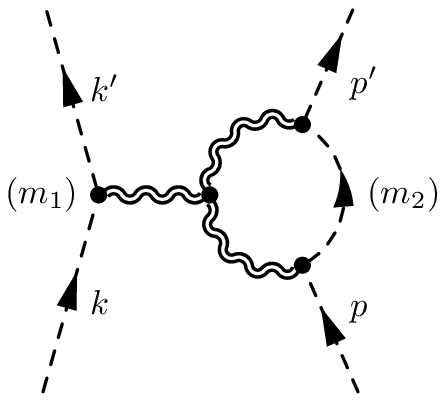}

\centering{(a)}
\end{minipage}
\begin{minipage}{0.45\linewidth}
\includegraphics[scale=.75]{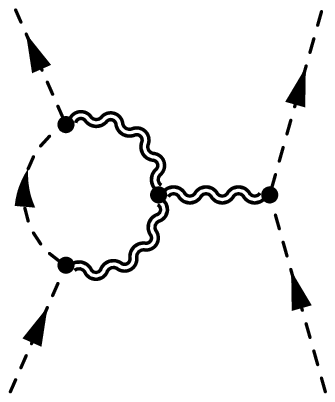}

\centering{(b)}
\end{minipage}
\begin{minipage}{0.45\linewidth}
\includegraphics[scale=.75]{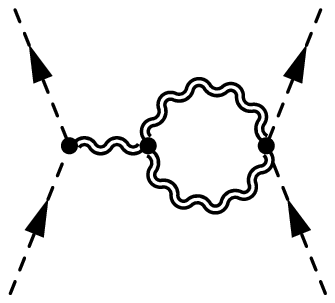}

\centering{(c)}
\end{minipage}
\begin{minipage}{0.45\linewidth}
\includegraphics[scale=.75]{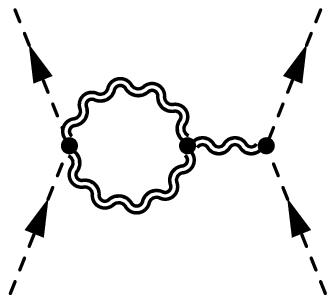}

\centering{(d)}
\end{minipage}

\caption{Vertex corrections from purely gravitational effects. (a) and (b) contain the $\phi^2 \, h$ coupling. 
(c) and (d) the $\phi^2 \, h^2$ coupling.}\label{VerCor}
\end{figure}

\begin{figure}[b]
\begin{minipage}{0.45\linewidth}
\includegraphics[scale=.75]{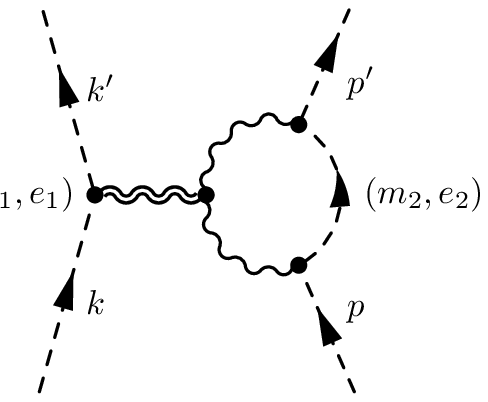}

\centering{(a)}
\end{minipage}
\begin{minipage}{0.45\linewidth}
\includegraphics[scale=.75]{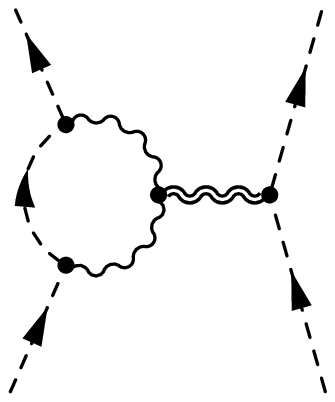}

\centering{(b)}
\end{minipage}
\begin{minipage}{0.45\linewidth}
\includegraphics[scale=.75]{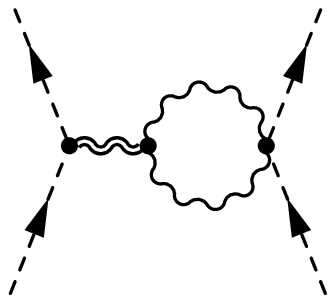}

\centering{(c)}
\end{minipage}
\begin{minipage}{0.45\linewidth}
 \includegraphics[scale=.75]{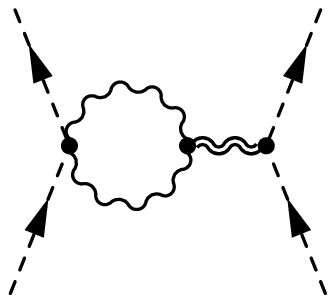}

\centering{(d)}
\end{minipage}
\caption{Vertex corrections involving gravitational and electromagnetic interactions (graviton 1-particle reducible).}\label{VerCorS1}
\end{figure}

\begin{figure}[t]
\begin{minipage}{0.45\linewidth}
\includegraphics[scale=.75]{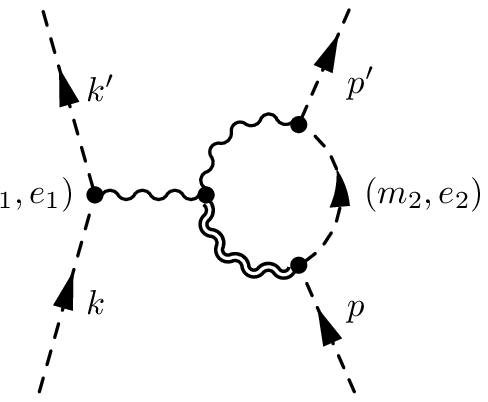}
\centering{(a)}
\end{minipage}
\begin{minipage}{0.45\linewidth}
\includegraphics[scale=.75]{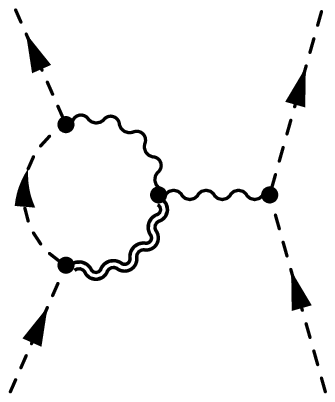}

\centering{(b)}
\end{minipage}
\begin{minipage}{0.45\linewidth}
\includegraphics[scale=.75]{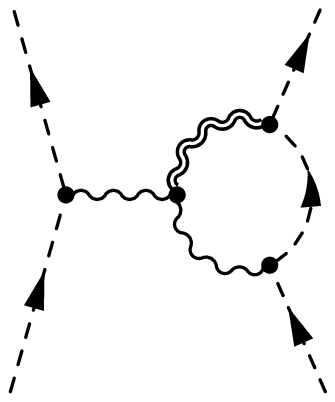}

\centering{(c)}
\end{minipage}
\begin{minipage}{0.45\linewidth}
\includegraphics[scale=.75]{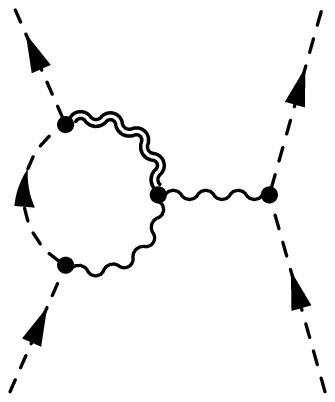}

\centering{(d)}
\end{minipage}
\begin{minipage}{0.45\linewidth}
\includegraphics[scale=.75]{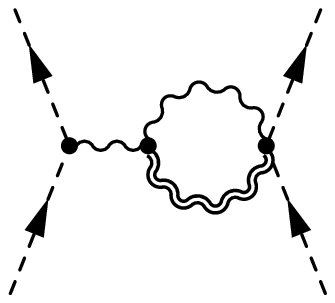}

\centering{(e)}
\end{minipage}
\begin{minipage}{0.45\linewidth}
 \includegraphics[scale=.75]{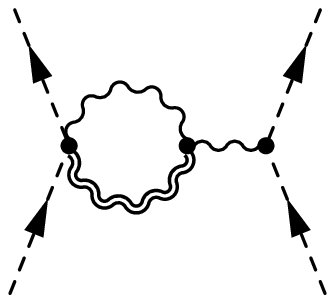}

\centering{(f)}
\end{minipage}

\caption{Vertex corrections involving gravitational and electromagnetic interactions (photon 1-particle reducible). }\label{VerCorS2}
\end{figure}

There are two classes of vertex corrections, involving gravitational effects only,  as shown in Fig.~\ref{VerCor}. 
The  contributions from Fig.~\ref{VerCor}(a) have the form
\begin{align}
i \M_{\ref{VerCor}(a)}(q) = &\tau_1^{\mu\nu} (k,k',m_1) \biggl \lbrack \frac{i  \mathcal P_{\mu\nu\alpha\beta}}{q^2}\biggr \rbrack  \int \frac{d^4 \ell}{(2\pi)^4} \notag \\
	&\cdot \tau^{\alpha\beta\rho\sigma \gamma\delta}_{7}(\ell, -q)\biggl \lbrack \frac{i \mathcal P_{\rho\sigma\lambda\epsilon}}{\ell^2}\biggr \rbrack \tau_1^{\lambda\epsilon}(p,p-\ell,m_2) \notag \\
	&\cdot  \biggl \lbrack \frac{i }{(\ell - p)^2 - m_2^2} \biggr \rbrack \tau^{\tau\phi}_{1} (p-\ell, p',m_2)  \notag\\
	&\cdot \biggl \lbrack \frac{i \mathcal P_{\gamma\delta\tau\phi}}{(\ell + q)^2}\biggr \rbrack \; ,\label{eq:24}
\end{align}
and for Fig.~\ref{VerCor}(b) 
\begin{align}
i \M_{\ref{VerCor}(b)}(q) =&\tau_1^{\mu\nu} (p,p',m_2) \biggl \lbrack \frac{i \mathcal P_{\alpha\beta\mu\nu}}{q^2} \biggr \rbrack  \int \frac{d^4 \ell}{(2\pi)^2} \notag \\
	&\cdot \tau_1^{\lambda\epsilon}(k,k+\ell,m_1)  \biggl \lbrack \frac{i \mathcal P_{\lambda\epsilon\rho\sigma}}{\ell^2}\biggr \rbrack   \tau_7^{\alpha\beta\rho\sigma\gamma\delta}(-\ell, q) \notag \\
	&\cdot  \biggl \lbrack \frac{i \mathcal P_{\gamma\delta\tau\phi}}{(\ell + q)^2}\biggr \rbrack \tau_1^{\tau\phi}(\ell + k, k',m_1) \notag\\
	&\cdot \biggl \lbrack \frac{i}{(\ell + k)^2 - m_1^2}\biggr \rbrack \; .\label{eq:25}
\end{align}
For the graviton loop diagrams, Fig.~\ref{VerCor}(c) and (d), we find
\begin{align}
 i \M_{\ref{VerCor}(c)}(q) =& \tau_1^{\mu\nu} (k,k',m_1)  \biggl \lbrack \frac{i \mathcal P_{\mu\nu\alpha\beta}}{q^2}\biggr \rbrack\frac{1}{2!} \int \frac{d^4 \ell}{(2 \pi)^4} \notag \\
	&\cdot  \tau_7^{\alpha\beta\rho\sigma\gamma\delta}(\ell, -q)  \biggl \lbrack \frac{i \mathcal P_{\rho\sigma\lambda\epsilon}}{\ell^2} \biggr\rbrack \tau_3^{\tau\phi\lambda\epsilon}(p, p',m_2) \notag \\
	&\cdot \biggl \lbrack \frac{i \mathcal P_{\tau\phi\gamma\delta}}{(\ell + q)^2} \biggr \rbrack \; ,\label{eq:26}\allowdisplaybreaks\\
i \M_{\ref{VerCor}(d)}(q) =&\tau_1^{\mu\nu} (p,p',m_2) \biggl \lbrack \frac{i \mathcal P_{\alpha\beta\mu\nu}}{q^2}\biggr \rbrack\frac{1}{2!}  \int \frac{d^4 \ell}{(2 \pi)^2} \notag \\
 	&\cdot \tau_3^{\tau\phi\lambda\epsilon} (k,k',m_1) \biggl \lbrack \frac{i \mathcal P_{\lambda \epsilon \rho\sigma}}{\ell^2}\biggr \rbrack \tau_7^{\alpha\beta\rho\sigma\gamma\delta}(-\ell,q) \notag \\
	&\cdot  \biggl \lbrack \frac{i \mathcal P_{\gamma\delta\tau\phi}}{(\ell +q)^2}\biggr \rbrack \; .\label{eq:27} 
\end{align} 
Again, the various functions $\tau_i$ can be found in the Appendix.
Contracting the Lorentz indices and performing the loop integration, we find
 \begin{align}
i &\M_{\ref{VerCor}(a+b)}(q) = \notag \\
	& - 8\, i \, G^2 m_1^2 m_2^2 \biggl \lbrack \frac{\pi^2 (m_1 + m_2)}{\sqrt{-q^2}}+\frac{5}{3} \ \log (-q^2) \biggr \rbrack \; ,\label{eq:28} \\
i &\M_{\ref{VerCor}(c+d)}(q)= \frac{208 \, i}{3} \ G^2 m_1^2 m_2^2 \ \log (-q^2)\label{eq:29} \,.
 \end{align}
The Fourier transform  yields a contribution to the nonrelativistic potential
\begin{align}
V_{\ref{VerCor}(a+b)} (r) =& \frac{G^2 m_1 m_2 (m_1 + m_2)}{r^2} \notag \\
	& - \frac{5}{3}\frac{G^2 m_1 m_2}{\pi r^3}\; , \label{eq: 30}\\
V_{\ref{VerCor}(c+d)} (r) =& \frac{26}{3} \frac{G^2 m_1 m_2}{\pi r^3} \; .\label{eq:31}
\end{align}
Including electromagnetic interactions in SQED we obtain additional
vertex corrections, shown in Figs.~\ref{VerCorS1}, \ref{VerCorS2}. 
The results for the scattering amplitudes from the diagrams in Figs.~\ref{VerCorS1}(a) -- (d) are given by
\begin{align}
i\M_{\ref{VerCorS1}(a+b)}(q) =& - \frac{5 \, i}{2} \ G  \bigl( e_1^2 m_2^2 + e_2^2 m_1^2 \bigr) \log (-q^2) \notag \\
	&- i \, G \bigl (e_2^2 m_1^2 m_2 +e_1^2 m_2^2 m_1 \bigr ) \frac{\pi}{\sqrt{-q^2}} \; ,\label{eq:32}\\
i \M_{\ref{VerCorS1}(c+d)}(q)&= - \frac{i}{6} \ G\bigl( e_1^2 m_2^2 + e_2^2 m_1^2) \ \log (-q^2) \; ,\label{eq:33}
\end{align}
which contributes to the nonrelativistic potential as
\begin{equation}\label{eq:34}
\begin{split}
 V_{\ref{VerCorS1}(a-d)} (r) = &\frac{G \bigl( m_1 e_2^2 + m_2 e_1^2 \bigr)}{8 \pi^2 r^2} \\
	& - \frac{G \bigl(\frac{m_2}{m_1}e_1^2  + \frac{m_1}{m_2}e_2^2 \bigr)}{3 \pi r^3} \; .
\end{split}
\end{equation}
Note, that the diagrams Figs. \ref{VerCorS1}(c) and (d) are associated with a symmetry factor $1/2$. 
Further details of the computation can be found in Appendix \ref{A-D}.

Similarly, the  diagrams in Fig.~\ref{VerCorS2} yield
\begin{equation}\label{eq:35}
 i \M_{\ref{VerCorS2}(a-f)}(q) = 2\, i\, G\, m_1 m_2 e_1 e_2 \ \frac{\pi (m_1 + m_2)}{\sqrt{-q^2}} \; ,
\end{equation}
and
\begin{equation}\label{eq:36}
 V_{\ref{VerCorS2}(a-f)} (r) = - \frac{G \, e_1 e_2 (m_1 + m_2)}{4 \pi r^2} \; .
\end{equation}
%

\subsection{Triangle Diagrams}

\begin{figure}[t!!]
 \begin{minipage}{0.45\linewidth}
\includegraphics[scale=.75]{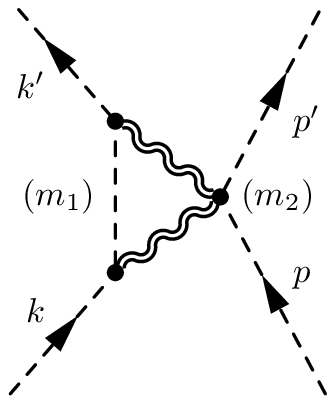}

\centering{(a)}
\end{minipage}
\begin{minipage}{0.45\linewidth}
\includegraphics[scale=.75]{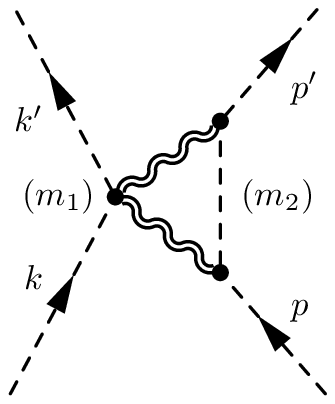}

\centering{(b)}
\end{minipage}

\caption{Triangle diagrams with only gravitational interactions.}\label{Tri}
\end{figure}
\begin{figure}[t]
\begin{minipage}{0.45\linewidth}
\includegraphics[scale=.75]{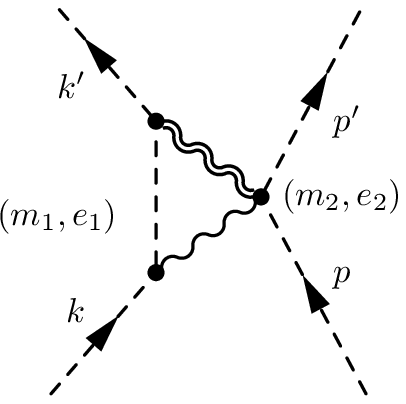}

\centering{(a)}
\end{minipage}
\begin{minipage}{0.45\linewidth}
\includegraphics[scale=.75]{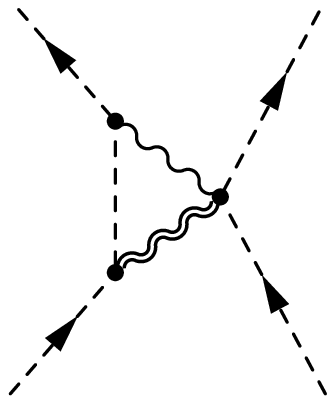}

\centering{(b)}
\end{minipage}
\begin{minipage}{0.45\linewidth}
\includegraphics[scale=.75]{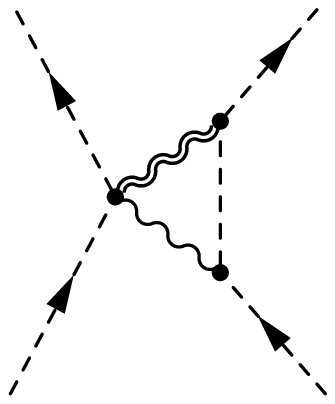}

\centering{(c)}
\end{minipage}
\begin{minipage}{0.45\linewidth}
\includegraphics[scale=.75]{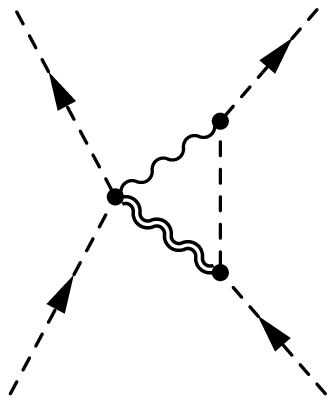}

\centering{(d)}
\end{minipage}

\caption{Triangle diagrams with gravitational and electromagnetic interactions.}\label{TriS}
\end{figure}

The triangle diagrams in Fig.~\ref{Tri} arise from purely gravitational interactions.
From the expressions in Appendix \ref{A-D} we compute
\begin{equation}\label{eq:37}
\begin{split}
 i \M_{\ref{Tri}(a+b)}(q) =& 8\, i\, G^2 m_1^2 m_2^2 \biggl \lbrack 28 \log (-q^2) \\
	&\phantom{ 8 i G^2 m_1^2 m_2^2} + \frac{4 \pi^2 (m_1 + m_2)}{\sqrt{-q^2}}\biggr \rbrack
\end{split}
\end{equation}
which contributes to the potential as
\begin{equation}\label{eq:38}
 V_{\ref{Tri}(a+b)} (r) = - \frac{G^2 m_1m_2}{r}\biggl \lbrack \frac{4(m_1+m_2)}{r} - \frac{28}{\pi r^2}\biggr \rbrack \; .
\end{equation}
The additional triangle diagrams in SQED are shown in
Fig.~\ref{TriS}. 
Their contribution to the scattering amplitude reads
\begin{align}
 i \M_{\ref{TriS}(a-d)}(q) = &4\, i \, m_1 m_2 \frac{G\, e_1 e_2}{\pi} \biggl \lbrack - 8 \log (-q^2) \notag \\
	&\phantom{4 i m_1 m_2}- \frac{2\pi^2 (m_1 +m_2)}{\sqrt{-q^2}}\biggr \rbrack \; ,\label{eq:39}
\end{align}
resulting in the potential
\begin{equation}\label{eq:40}
V_{\ref{TriS}(a-d)}(r) = \frac{G\, e_1 e_2 (m_1 + m_2)}{\pi r^2} - \frac{4\, G\,  e_1 e_2}{\pi^2 r^3} \; .
\end{equation}
%

\subsection{Box and Crossed Box Diagrams}
\begin{figure} [pt]
\begin{minipage}{0.45\linewidth}
\includegraphics[scale=.75]{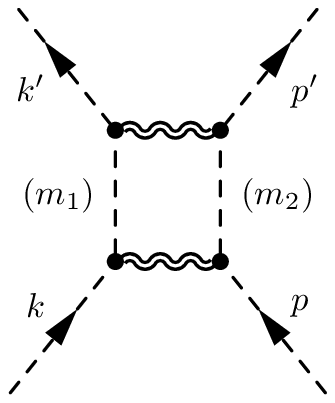}

\centering{(a)}
\end{minipage}
\begin{minipage}{0.45\linewidth}
\includegraphics[scale=.75]{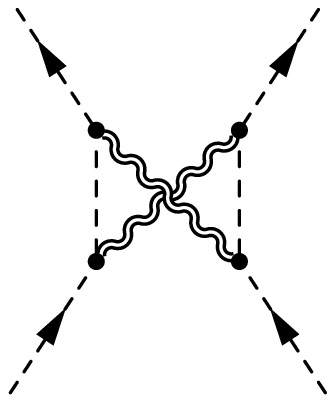}

\centering{(b)}
\end{minipage}

\caption{Box and crossed box diagrams from purely gravitational interactions.}\label{BC}
\end{figure}
The contributions of the box and crossed box diagrams in Fig.~\ref{BC} read
\begin{align}
 i \M_{\ref{BC}(a)} (q) =& \int \frac{d^4 \ell}{(2\pi)^4} \tau_1^{\mu\nu}(k,k+\ell,m_1) \biggl \lbrack \frac{i \mathcal P_{\mu\nu\alpha\beta}}{\ell^2}\biggr \rbrack \notag\\
	&\cdot \tau_1^{\alpha\beta}(p,p-\ell,m_2) \biggl \lbrack \frac{i}{(\ell - p)^2 -m_2^2}\biggr \rbrack \notag \\
	&\cdot \tau_1^{\gamma\delta}(p-\ell, p',m_2) \biggl \lbrack\frac{i \mathcal P_{\gamma\delta\rho\sigma}}{(\ell +q)^2}\biggr \rbrack\notag\\
	&\cdot \tau_1^{\rho\sigma}(k+\ell, k',m_1) \biggl \lbrack \frac{i}{(\ell + k)^2 - m_1^2}\biggr \rbrack \; ,\label{eq:41}
\end{align}
and 
\begin{align}
i \M_{\ref{BC}(b)} (q)  =& \int \frac{d^4\ell}{(2\pi)^4} \tau_1^{\mu\nu}(k, \ell + k,m_1) \biggl \lbrack \frac{i \mathcal P_{\mu\nu\alpha\beta}}{\ell^2}\biggr \rbrack \notag \\
	&\cdot \tau_1^{\alpha\beta}(\ell + p',p',m_2)\biggl \lbrack \frac{i}{(\ell + p')^2 - m_2^2}\biggr \rbrack \notag \\
	&\cdot \tau_1^{\gamma\delta}(p,\ell +p',m_2) \biggl \lbrack \frac{i \mathcal P_{\gamma\delta\rho\sigma}}{(\ell + q)^2}\biggr \rbrack\notag \\
	&\cdot \tau_1^{\rho\sigma}(k+\ell , k', m_1)\biggl \lbrack \frac{i}{(\ell + k)^2 -m_1^2}\biggr \rbrack \; .\label{eq:42}
\end{align}
Again, we are only interested in the nonanalytic terms, and one may exploit some
simplifications which reduce parts of the amplitude to three- and two-point functions,
see Appendix \ref{A-B},
\begin{equation}\label{eq:43}
 i \M_{\ref{BC}(a+b)}^\text{red}(q)  = - 64 \, i \, G^2 m_1^2 m_2^2 \, \log (-q^2) \; .
\end{equation}
For the remaining irreducible parts we find
\begin{equation}\label{eq:44}
 i \M_{\ref{BC}(a+b)}^\text{irred}(q) = - \frac{184}{3} \,  i  \, G^2  m_1^2 m_2^2 \, \log (-q^2) \,.
\end{equation}
In the nonrelativistic limit this combines to the  potential
\begin{equation}\label{eq:45}
V_{\ref{BC}(a+b)}= - \frac{47}{3} \frac{G^2 \, m_1 m_2}{\pi r^3} \; .
\end{equation}
\begin{figure}[t]
\begin{minipage}{0.45\linewidth}
\includegraphics[scale=.75]{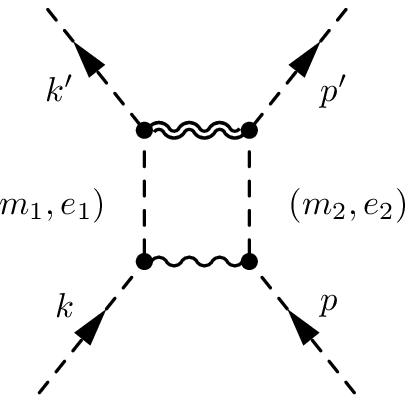}

\centering{(a)}
\end{minipage}
\begin{minipage}{0.45\linewidth}
\includegraphics[scale=.75]{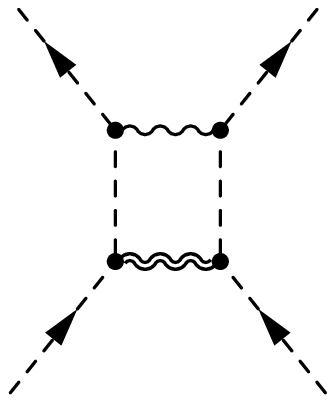}

\centering{(b)}
\end{minipage}
\begin{minipage}{0.45\linewidth}
\includegraphics[scale=.75]{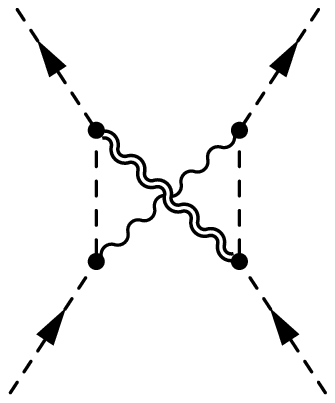}

\centering{(c)}
\end{minipage}
\begin{minipage}{0.45\linewidth}
\includegraphics[scale=.75]{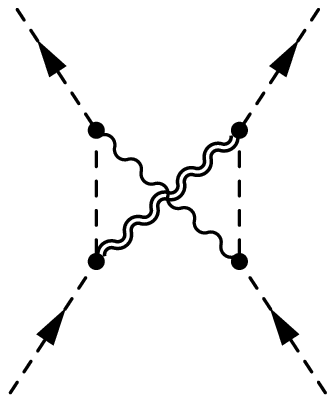}

\centering{(d)}
\end{minipage}
\caption{Box and crossed box diagrams with gravitational and electromagnetic interactions.}\label{BCS}
\end{figure}
Similarly, including electromagnetic interactions,
the box and crossed box diagrams in Fig.~\ref{BCS} give a
reducible part,
\begin{equation}\label{eq:46}
 i \M_{\ref{BCS}(a-d)}^\text{red}(q)= 32\, i \, G \, m_1 m_2 \  \frac{e_1 e_2}{4\pi}  \ \log (-q^2) \; ,
\end{equation}
and an irreducible part,
\begin{equation}\label{eq:47}
 i \M_{\ref{BCS}(a-d)}^\text{irred} (q)= \frac{224 \, i}{3} \, G \, m_1 m_2 \, \frac{e_1 e_2}{4\pi} \, \log (-q^2)
\end{equation}
The contribution to the potential is
\begin{equation}\label{eq:48}
 V_{\ref{BCS}(a-d)} (r) = \frac{10}{3} \ \frac{G \, e_1 e_2}{\pi^2 r^3} \; .
\end{equation}
%

\subsection{Circular Diagram}

\begin{figure}[b]
\begin{minipage}{0.45\linewidth}
\includegraphics[scale=.75]{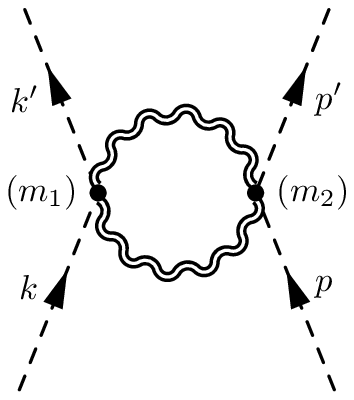}

\centering{(a)}
\end{minipage}
\begin{minipage}{0.45\linewidth}
\includegraphics[scale=.75]{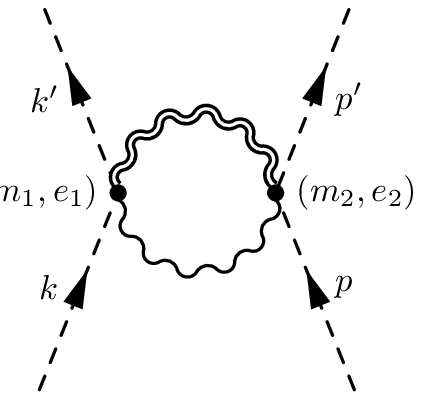}

\centering{(b)}
\end{minipage}

\caption{Circular diagrams: (a) purely gravitational; (b) gravitational and electromagnetic interactions .}\label{CD}
\end{figure}
The calculation of the  circular diagrams in Fig.~\ref{CD} is straightforward. We find 
the nonanalytic contributions to the scattering amplitude
\begin{align}
i \M_{\ref{CD}(a)}(q) =& - 176 \ i \ G^2 \ m_1^2 m_2^2 \ \log (-q^2)\label{eq:49} \\
\intertext{and}
i \M_{\ref{CD}(b)}(q) =& \frac{16 \,  i}{\pi} \, G  m_1 m_2 \, e_1 e_2 \, \log (-q^2)  \label{eq:50} \,,
\end{align}
and the resulting contribution to the potential
\begin{align}
 V_{\ref{CD}(a)} (r) =& -22 \,  \frac{G^2 \, m_1 m_2}{\pi r^3} \label{eq:51}\,, \\
\intertext{and}
V_{\ref{CD}(b)}(r) =&\, 2 \,  \frac{G \, e_1 e_2}{\pi^2 r^3} \; . \label{eq:52} 
\end{align}
%

\subsection{Vacuum Polarization Diagram}

\begin{figure}[bpt]
 \begin{minipage}{0.45\linewidth}
\includegraphics[scale=.75]{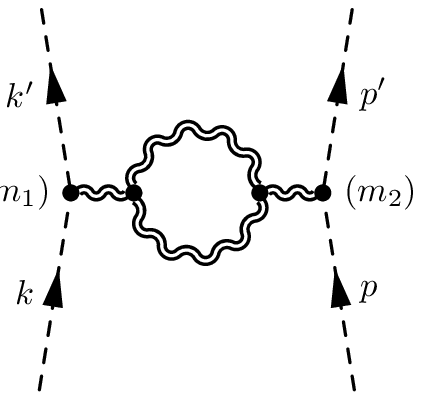}

\centering{(a)}
\end{minipage}
\begin{minipage}{0.45\linewidth}
\includegraphics[scale=.75]{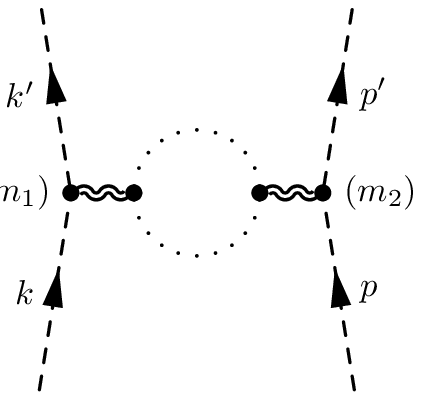}

\centering{(b)}
\end{minipage}
\begin{minipage}{0.45\linewidth}
\includegraphics[scale=.75]{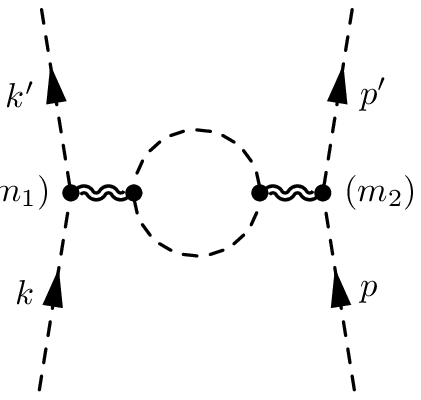}

\centering{(c)}
\end{minipage}
\begin{minipage}{0.45\linewidth}
\includegraphics[scale=.75]{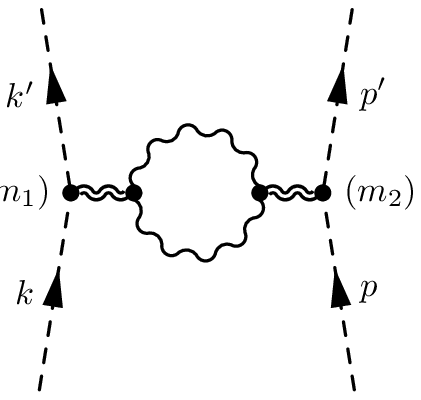}

\centering{(d)}
\end{minipage}
\caption{The set of vacuum polarization diagrams which contribute to the potential.}\label{VacP}
\end{figure}

\begin{figure}[t!!]
\begin{minipage}{0.30\linewidth}
\vspace{2.5mm}
\includegraphics[scale=.75]{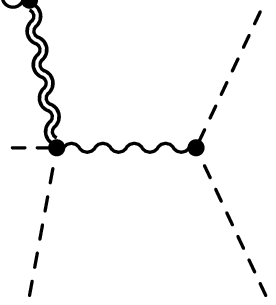}

\centering{(a)}
\end{minipage}
\begin{minipage}{0.30\linewidth}
\includegraphics[scale=.75]{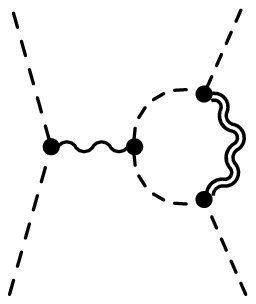}

\centering{(b)}
\end{minipage}
\begin{minipage}{0.30\linewidth}
\includegraphics[scale=.75]{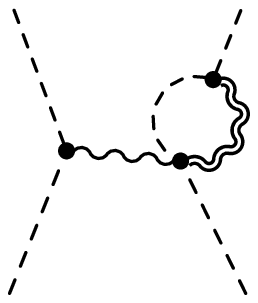}

\centering{(c)}
\end{minipage}

\caption{The set of diagrams which only give analytical contributions to the scattering amplitude.}\label{PAC}
\end{figure}

Treating general relativity as an effective field theory, we have to deal with the
vacuum polarization diagrams Fig.~\ref{VacP}(a)--(d) as well.
The graviton self-energy is obtained from the subdiagrams in Fig.~\ref{VacP}(a-b). Its contribution to
the effective Lagrangian has been worked out
by 't~Hooft and Veltman \cite{Hoo74, Vel76, Don94a, Don94b},
\begin{equation}\label{eq:53}
 \La =  \frac{-1}{16\pi^2} \biggl ( \frac{1}{120} \ \bar{\mathscr R}^2 + \frac{7}{20} \ \mathcal R_{\mu\nu}\mathcal R^{\mu\nu}\biggr ) \ \log (-q^2) \; .
\end{equation}
From this Lagrangian, one obtains the vacuum polarization tensor
\begin{align}
 \tilde\Pi _{\alpha\beta, \gamma\delta} =  \frac{-2\, G}{\pi } \biggl \lbrack &\frac{21}{120} q^4 \mathds 1_{\alpha\beta\gamma\delta}+\frac{23}{120}q^4 \eta_{\alpha\beta}\eta_{\gamma\delta} \notag \\
	&- \frac{23}{120} \bigl( \eta_{\alpha\beta}q_\gamma q_\delta + \eta_{\gamma\delta}q_\alpha q_\beta \bigr) q^2  \notag \\
	&- \frac{21}{240}\bigl( q_\alpha q_\delta \eta_{\beta\gamma} + q_\beta q_\delta \eta_{\alpha\gamma}\notag \\
	&+ q_\alpha q_\gamma \eta_{\beta\delta} + q_\beta q_\gamma \eta_{\alpha\delta}\bigr) q^2 \notag \\
	&+ \frac{11}{30} q_\alpha q_\beta q_\gamma q_\delta \biggr \rbrack \ \log (-q^2) \; .\label{eq:54}
\end{align}
Inserting this into the scattering amplitudes in Fig.~\ref{VacP}(a+b), one obtains
\begin{align}
 i \M_{\ref{VacP}(a+b)} (q) = &\tau_1^{\mu\nu} (k,k',m_1)\biggl \lbrack \frac{i \mathcal P_{\mu\nu\alpha\beta}}{q^2}\biggr \rbrack \notag \\
	&\cdot \tilde\Pi^{\alpha\beta, \gamma\delta} (q)\biggl \lbrack \frac{i \mathcal P_{\gamma\delta \rho \sigma}}{q^2}\biggr \rbrack \tau_1^{\rho\sigma}(p,p',m_2) \; ,	
\notag \\
	= & - \frac{172\, i}{15} \ G^2 m_1^2 m_2^2 \ \log (-q^2) \,, \label{eq:55}
\end{align}
which yields the potential
\begin{equation}\label{eq:56}
 V_{\ref{VacP}(a+b)} (r) = - \frac{43}{30} \ \frac{G^2 \, m_1 m_2}{\pi r^3} \; .
\end{equation}
Furthermore we have to treat the scalar and photon contribution to the graviton
self-energy in Figs.~\ref{VacP}(c) and \ref{VacP}(d), respectively.
The contribution of massive scalar particles only contains analytical terms
and is therefore irrelevant for the calculation of the nonrelativistic potential.
If we have $N$ additional massless (uncharged) scalars in the theory, they would contribute
to the graviton polarization tensor as 
\begin{align}
\bar\Pi_{\alpha\beta, \gamma\delta} =& N \, \frac{- G}{\pi} \biggl \lbrack \frac{1}{20} \bigl(q_\alpha q_\beta - q^2 \eta_{\alpha\beta}\bigr)  \bigr( q_\gamma q_\delta - q^2 \eta_{\gamma\delta}\bigr) \notag \\
	&+ \frac{1}{30} \bigl( q_\alpha q_\gamma -q^2 \eta_{\alpha\gamma} \bigr) \bigl( q_\beta q_\delta - q^2 \eta_{\beta\delta}\bigr)\notag \\
	&+ \frac{1}{30} (q_\alpha q_\delta - q^2 \eta_{\alpha\delta}\bigr) \bigl( q_\beta q_\gamma -q^2 \eta_{\beta\gamma}\bigr) \biggr \rbrack \log (-q^2) \,.\label{eq:57}
\end{align}
Inserting this into the scattering amplitude yields
\begin{equation}\label{eq:58}
 i \M_{\ref{VacP}(c)}(q) = - N \, \frac{8 \, i}{5} \ G^2\, m_1^2 m_2^2 \ \log (-q^2) \,,
\end{equation}
which corresponds to a contribution to the nonrelativistic potential
\begin{equation}\label{eq:59}
 V_{\ref{VacP}(c)}(r) = \frac{-N}{5} \ \frac{G^2\, m_1 m_2 }{ \pi r^3} \; .
\end{equation}
Finally, one calculates the photon-loop contribution to the polarization tensor
\begin{align}
 \hat\Pi_{\alpha\beta, \gamma\delta} =& \frac{G}{\pi} \biggl \lbrack \frac{1}{15}\bigl(q_\alpha q_\beta - q^2 \eta_{\alpha\beta} \bigr)\bigl( q_\gamma q_\delta - q^2 \eta_{\gamma \delta}\bigr) \notag \\
	& - \frac{1}{10} \bigl( q_\alpha q_\gamma - q^2 \eta_{\alpha\gamma}\bigr) \bigl( q_\beta q_\delta - q^2 \eta_{\beta\delta}\bigr) \notag \\
	&- \frac{1}{10} \bigl( q_\alpha q_\delta - q^2 \eta_{\alpha\delta}\bigr)\bigl( q_\beta q_\gamma - q^2 \eta_{\beta\gamma}\bigr) \biggr \rbrack \ \log (-q^2) \; ,\label{eq:60}
\end{align}
which yields
\begin{align}
 i \M_{\ref{VacP}(d)} (q) =& -  \frac{64\, i}{15} \ G^2 m_1^2 m_2^2 \ \log (-q^2) \label{eq:61}\,, \\
\intertext{and}
V_{\ref{VacP}(d)}=& - \frac{4}{15}\ \frac{G^2\, m_1 m_2}{\pi r^2}\label{eq:62} \,.
\end{align}
\begin{figure}[bpt]
\begin{minipage}{0.45\linewidth}
\includegraphics[scale=.75]{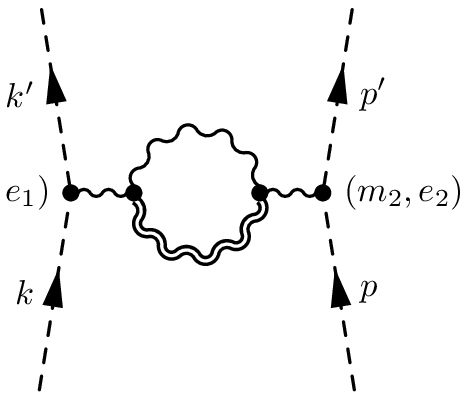}

\end{minipage}

\caption{The one-loop contribution to the photon self-energy in SQED.}\label{VacPS}
\end{figure}
The photon self-energy in SQED only receives a one-loop contribution
from the mixed vacuum polarization diagram in Fig.~\ref{VacPS},
\begin{align} 
\Pi^{\gamma\tau} (q) =& \frac{i \, G}{\pi} \ \biggl( \frac{q^2}{3}\biggr) \bigl( q^\gamma q^\tau - q^2 \eta^{\gamma\tau} \bigr) \ \log (-q^2) \; ,\label{eq:63}
\end{align}
which yields
\begin{align}
 i \M_{\ref{VacPS}}(q) =&\frac{4\, i }{3 \pi} \, G \, m_1 m_2 \,  e_1 e_2 \ \log (-q^2)  \label{eq:64} \,, \\
\intertext{and}
V_{\ref{VacPS}} (r) =&  \frac{1}{6} \frac{G \, e_1 e_2}{\pi^2 r^3}\label{eq:65} \,.
\end{align}
We note that the one-loop diagrams in Fig.~\ref{PAC} only contribute analytical terms
to the scattering amplitude and thus do not contribute to the nonrelativistic potential,
see also \cite{Don94b,Bje02}. 
%

\subsection{Final result for the nonrelativistic potential}

In this paragraph we are going to summarize the main result of our paper:
the nonrelativistic potential which is obtained from combining the contributions of the various
diagrams discussed above. 
Let us first concentrate on the purely gravitational
effects, for which we obtain
\begin{align}\
 V_\text{GR}(r) = - \frac{G \, m_1 m_2}{r} \biggl \lbrack 1&+ 3 \, \frac{G \, (m_1 + m_2)}{\text c^2 r} \notag \\
	& + \frac{131+6N}{30 \pi} \, \frac{G  \hbar}{\text c^3 r^2}\biggr \rbrack \; .\label{eq:66}
\end{align}
Here, we have restored the dependence on $c$ and $\hbar$ to distinguish the effects from general relativity
and quantum corrections. We recall that $N$ denotes the number of massless (uncharged) scalars, e.g.,\
``scalar neutrinos.''  When comparing our result (\ref{eq:66}) with \cite{Bje03}, we have to take
into account that in \cite{Bje03} the quantum corrections to the graviton propagator from massless scalar
and photon loops have not been considered. Apart from this, we fully agree with  \cite{Bje03}. 
The discrepancies between  \cite{Bje01,Bje03} and 
the original calculations in  \cite{Don94a, Don94b, Don95} and the results of \cite{Akh97} have already been
discussed, and we thus confirm the conclusions of  \cite{Bje03}.
Concerning the contributions from massless scalars to the polarization tensor in (\ref{eq:57}),
we disagree with a previous calculation in \cite{Ham95}. The photon contribution in (\ref{eq:60})
coincides with the result in \cite{Cap74}.

Including electromagnetic effects in scalar QED, we find
\begin{align}
 V_\text{SQED}(r) =& V_\text{GR}(r)  \nonumber \\
&
 + \frac{\tilde \alpha \mathfrak e_1 \mathfrak e_2}{r}
\left( 1+ 3 \ \frac{G \, (m_1 + m_2) }{\text c^2 r}
+\frac{6}{\pi} \,  \frac{ G \hbar}{\text c^3 r^2} \right) \nonumber \\
 &+ \frac{1}{2}\frac{(m_1 \mathfrak e_2^2 + m_2 \mathfrak e_1^2) \ G \tilde \alpha}{\text c^2 r^2} \notag \\
	& - \frac{4}{3\pi}\biggl( \frac{m_2^2 \mathfrak e_1^2 + m_1^2 \mathfrak e_2^2}{m_1 m_2}\biggr) \frac{G \tilde \alpha \hbar}{\text c^3 r^3} \; .\label{eq:67}
\end{align}
Here we have introduced $\tilde \alpha = \hbar \text c / 137$
and the charges $\mathfrak e_1$ and $\mathfrak e_2$ are normalized in units of the elementary charge.
Again, our result for the potential \eqref{eq:67} is in agreement
with \cite{Bje02,Hol08}. [A calculational error in the expression for the box diagrams (\ref{eq:47}) quoted in 
a previous preprint version of this paper \cite{Fa07}  has been corrected.] 
%

\section{Summary}

We have calculated the one-loop corrections to the Newtonian and Coulomb potential
for massive and electrically charged scalar particles,
treating general relativity as an effective quantum field theory.
The nonrelativistic potential is obtained from the scattering amplitude
by considering the nonanalytic terms in the momentum transfer $q^2$.
After Fourier transformation, 
we identify relativistic corrections of order  $Gm/\text c^2 r$ 
and genuine quantum corrections of order $G\hbar/\text c^3 r^2$. 
Our results for the leading quantum corrections to the Newtonian and Coulomb laws
are in agreement with \cite{Bje03} and \cite{Bje02,Hol08}.
Concerning the quantum corrections to the graviton propagator,
we reproduced the contribution from the photon loop in \cite{Cap74},
but differ for the case of massless scalar loops in \cite{Ham95}.

It is straight forward to generalize the effective-theory formalism to, for instance,
the case of massive charged spin-$\frac{1}{2}$ fermions which
has been discussed in \cite{Bje03a, But06}. In this way one could 
unambiguously predict the quantum-gravitational corrections to low-energy
observables within the Standard Model.
However, one has to be aware that the size of the calculated corrections
is -- of course -- extremely small, $G\hbar/\text c^3 r^2 = 2.6 \cdot 10^{-40} \, {\rm fm^2}/r^2 $,
 and it will thus be practically impossible
to verify the theoretical results by experimental measurements in the
foreseeable future \cite{Don95}. 
%

\subsection*{Acknowledgments}

I would like to thank my supervisors Th.~Mannel and A.~Khodjamirian for the encouragement and
many helpful comments. 
Special thanks go to N. E. J. Bjerrum-Bohr, University
of Wales, Swansea, and Institute for Advanced Study, Princeton, 2006-2007,
for discussions on this topic, and to Th.~Feldmann for critical discussions and 
comments on this manuscript.
%

 \appendix

\section{Feynman Rules}\label{A-A}
%

\subsection{Propagators}
%

The propagator for a massive scalar field is given by
 \begin{align*}
  \parbox{60pt}{\includegraphics[scale=.75]{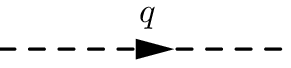}}\qquad &=  \frac{i}{q^2 - m^2 + i \epsilon}  \; .
 \end{align*}
%

The photon propagator in harmonic (Feynman) gauge is given by
 \begin{align*}
  \parbox{80pt}{\includegraphics[scale=.75]{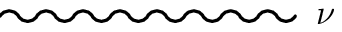}}\qquad &=  \frac{-i \eta^{\mu\nu}}{q^2 + i \epsilon} \; .
 \end{align*}
%

In harmonic gauge the graviton propagator has been discussed in \cite{Don94b,Hoo74}. 
An explicit derivation can be found in \cite{Bje01} with the result
\begin{align*}
 \parbox{100pt}{\includegraphics[scale=.75]{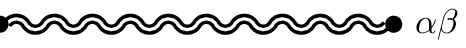}}\qquad&= \frac{i \mathcal P^{\mu\nu\alpha\beta}}{q^2 + i \epsilon}
\end{align*}
where the polarization sum is given by
\begin{equation*}
 \mathcal P^{\mu\nu\alpha\beta} = \frac{1}{2}\bigl( \eta^{\mu\alpha}\eta^{\nu\beta}+\eta^{\mu\beta}\eta^{\nu\alpha}- \eta^{\mu\nu}\eta^{\alpha\beta}\bigr) \; .
\end{equation*}
%

\subsection{Vertices}
%

The 1-graviton-2-scalar vertex is given by
\begin{align*}
 \parbox{60pt}{\includegraphics[scale=.75]{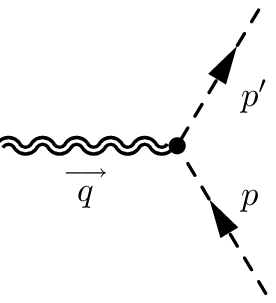}} \qquad &= \tau_1^{\mu\nu} (p,p',m) \ ,
\end{align*}
with 
\begin{equation*}
 \tau_1^{\mu\nu}(p,p', m) = \frac{-i \kappa}{2} \bigl \lbrace p^\mu p^{\prime \, \nu} + p^\nu p^{\prime \, \mu} - \eta^{\mu\nu} \bigl \lbrack ( p \cdot p' ) - m^2 \bigr \rbrack \bigr \rbrace
\end{equation*}
%

The 1-photon-2-scalar vertex is found to be
\begin{align*}
 \parbox{60pt}{\includegraphics[scale=.75]{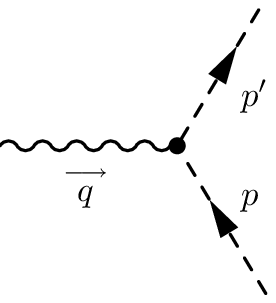}} \qquad &= \tau_2^{\gamma} (p, p',e)\\& = -i  e  \bigl(p^\gamma + p^{\prime \, \gamma} \bigr) \; .
\end{align*}
%

The 2-graviton-2-scalar-vertex reads
\begin{align*}
 \parbox{80pt}{\includegraphics[scale=.75]{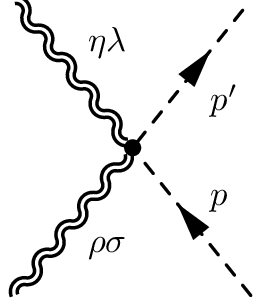}}\qquad &= \tau_3^{\eta\lambda\rho\sigma} (p,p',m)
\end{align*}
with
\begin{align*}
 \tau_3^{\eta\lambda\rho\sigma}& (p,p',m) = i \kappa^2\biggl \lbrace \biggl \lbrack \mathds 1^{\eta\lambda\alpha\delta}\tensor{\mathds 1}{^\rho^\sigma^\beta_\delta}- \frac{1}{4} \biggl( \eta^{\eta_\lambda}\mathds 1^{\rho\sigma\alpha\beta} \\
	&+ \eta^{\rho\sigma} \mathds 1^{\eta\lambda\alpha\beta}\biggr) \biggr \rbrack \bigl( p_\alpha p^\prime_\beta + p_\beta p^\prime_\alpha \bigr) - \frac{1}{2}\biggl \lbrack \mathds 1^{\eta\lambda\rho\sigma} \\
	&- \frac{1}{2} \eta^{\eta\lambda}\eta^{\rho\sigma} \biggr \rbrack \biggl( (p \cdot p') - m^2 \biggr) \biggr \rbrace
\end{align*}
and
\begin{equation*}
 \mathds 1_{\alpha\beta\gamma\delta} = \frac{1}{2}\bigl( \eta_{\alpha\gamma}\eta_{\beta\delta}+ \eta_{\alpha\delta}\eta_{\beta\gamma}\bigr) \ .
\end{equation*}
%

The 2-photon-1-graviton vertex is given by
\begin{align*}
 \parbox{80pt}{\includegraphics[scale=.75]{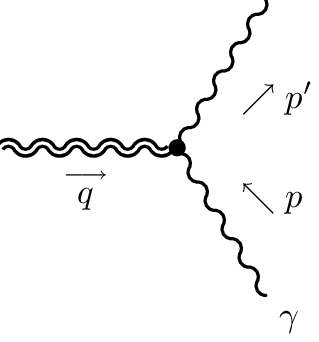}} &= \tau_4^{\mu\nu(\gamma\delta)} (p,p') \ ,
\end{align*}
with
\begin{align*}
 \tau_4^{\mu\nu(\gamma\delta)}&(p,p') = i \kappa \biggl \lbrace \mathcal P^{\mu\nu (\gamma \delta)} \ (p \cdot p') + \frac{1}{2} \biggl \lbrack \eta^{\mu\nu}p^\delta p^{\prime \, \gamma} \\
	&+ \eta^{\gamma\delta} \bigl( p^\mu p^{\prime \, \nu} + p^\nu p^{\prime \, \mu} \bigr) - \bigl (\eta^{\mu\delta} p^{\prime \, \gamma} p^\nu + \eta^{\nu\delta} p^{\prime \, \gamma} p^\mu \\
	&+\eta^{\nu\gamma}p^{\prime \,\mu}p^\delta + \eta^{\mu\gamma}p^{\prime \, \nu}p^\delta \bigr)\biggr \rbrack\biggr \rbrace  \ .
\end{align*}
%

The 2-scalar-2-photon vertex in scalar QED reads
\begin{align*}
 \parbox{80pt}{\includegraphics[scale=.75]{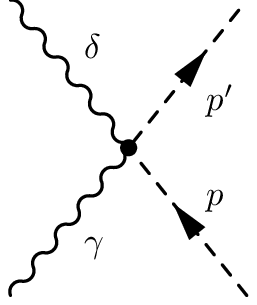}} &= \tau_5^{\gamma\delta} (p,p',e) = 2 \ i \ e^2 \ \eta^{\gamma\delta} \; .
\end{align*}
%

The 2-scalar-1-photon-1-graviton is found to be 
\begin{align*}
 \parbox{80pt}{\includegraphics[scale=.75]{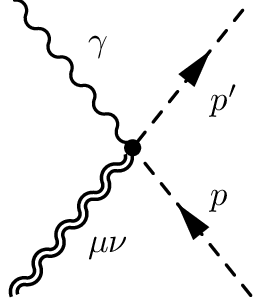}} &= \tau_6^{\mu\nu(\gamma)} (p,p',e) 
\end{align*}
with
\begin{equation*}
 \tau_6^{\mu\nu(\gamma)} (p,p',e) =  i \ e \ \kappa \bigl \lbrack \mathcal P^{\mu\nu \alpha\gamma} \bigl( p + p')_\alpha \bigr \rbrack\ .
\end{equation*}
%

Using the background field method the 3-graviton vertex is found as \cite{Don94a,Don94b,Bje02}
\begin{align*}
 \parbox{80pt}{\includegraphics[scale=.75]{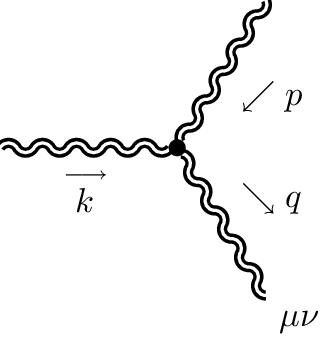}} &= \tau^{\mu\nu}_{7 \, \alpha\beta\gamma\delta} (k,q) \ ,
\end{align*}
with
\begin{widetext}
 \begin{equation*}
  \begin{split}
   \tau^{\mu\nu}_{7 \, \alpha\beta\gamma\delta} = -\frac{i \kappa}{2} &\cdot \biggl \lbrace \mathcal P_{\alpha\beta\gamma\delta} \biggl \lbrack k^\mu k^\nu + (k-q)^\mu  (k-q)^\nu + q^\mu q^\nu - \frac{3}{2} \eta^{\mu\nu} q^2 \biggr \rbrack + 2 q_\lambda q_\sigma \bigl \lbrack \tensor{\mathds 1}{_\alpha_\beta^\sigma^\lambda}\ \tensor{\mathds 1}{_\gamma_\delta^\mu^\nu} + \tensor{\mathds 1}{_\gamma_\delta^\sigma^\lambda} \ \tensor{\mathds 1}{_\alpha_\beta^\mu^\nu} - \tensor{\mathds 1}{_\alpha_\beta^\mu^\sigma} \ \tensor{\mathds 1}{_\gamma_\delta^\nu^\lambda} \\
	& - \tensor{\mathds 1}{_\gamma_\delta^\mu^\sigma} \ \tensor{\mathds 1}{_\alpha_\beta^\nu^\lambda} \bigr \rbrack+ \bigl \lbrack q_\lambda q^\mu \bigr( \eta_{\alpha\beta} \tensor{\mathds 1}{_\gamma_\delta^\nu^\lambda} + \eta_{\gamma\delta} \tensor{\mathds 1}{_\alpha_\beta^\nu^\lambda} \bigr) +q_\lambda q^\nu \bigr( \eta_{\alpha\beta} \tensor{\mathds 1}{_\gamma_\delta^\mu^\lambda} + \eta_{\gamma\delta} \tensor{\mathds 1}{_\alpha_\beta^\mu^\lambda} \bigr)\\
	&-q^2 \bigl( \eta_{\alpha\beta} \tensor{\mathds 1}{_\gamma_\delta^\mu^\nu}+ \eta_{\gamma\delta}\tensor{\mathds 1}{_\alpha_\beta^\mu^\nu}\bigr)- \eta^{\mu\nu} q_\lambda q_\sigma \bigr( \eta_{\alpha\beta}\tensor{\mathds 1}{_\gamma_\delta^\sigma^\lambda} + \eta_{\gamma\delta} \tensor{\mathds 1}{_\alpha_\beta^\sigma^\lambda} \bigr) \bigr \rbrack \\
	&+ \bigl \lbrack 2 q_\lambda \bigl \lbrace \tensor{\mathds 1}{_\alpha_\beta^\lambda^\sigma} \ \tensor{\mathds 1}{_\gamma_\delta_\sigma^\nu} (k-q)^\mu + \tensor{\mathds 1}{_\alpha_\beta^\lambda^\sigma} \ \tensor{\mathds 1}{_\gamma_\delta_\sigma^\mu} (k-q)^\nu - \tensor{\mathds 1}{_\gamma_\delta^\lambda^\sigma} \ \tensor{\mathds 1}{_\alpha_\beta_\sigma^\nu} k^\mu - \tensor{\mathds 1}{_\gamma_\delta^\lambda^\sigma} \ \tensor{\mathds 1}{_\alpha_\beta_\sigma^\mu} k^\nu \bigr \rbrace \\
	& + q^2 \bigl( \tensor{\mathds 1}{_\alpha_\beta_\sigma^\mu}\tensor{\mathds 1}{_\gamma_\delta^\nu^\sigma} + \tensor{\mathds 1}{_\alpha_\beta^\nu^\sigma}\tensor{\mathds 1}{_\gamma_\delta_\sigma^\mu} \bigr) + \eta^{\mu\nu} q_\sigma q_\lambda \bigl( \tensor{\mathds 1}{_\alpha_\beta^\lambda^\rho} \ \tensor{\mathds 1}{_\gamma_\delta_\rho^\sigma}  + \tensor{\mathds 1}{_\gamma_\delta^\lambda^\rho} \ \tensor{\mathds 1}{_\alpha_\beta_\rho^\sigma} \bigr) \bigr \rbrack + \pmb{\biggl \lbrace} \bigl( k^2 + (k-q)^2 \bigr) \\
	 &\cdot \biggl \lbrack \tensor{\mathds 1}{_\alpha_\beta^\mu^\sigma} \ \tensor{\mathds 1}{_\gamma_\delta_\sigma^\nu} +\tensor{\mathds 1}{_\gamma_\delta^\mu^\sigma} \ \tensor{\mathds 1}{_\alpha_\beta_\sigma^\nu} - \frac{1}{2} \eta^{\mu\nu} \mathcal P_{\alpha\beta\gamma\delta} \biggr \rbrack - \bigl( \tensor{\mathds 1}{_\gamma_\delta^\mu^\nu} \ \eta_{\alpha\beta} \ k^2 - \tensor{\mathds 1}{_\alpha_\beta^\mu^\nu} \ \eta_{\gamma\delta} \ (k-q)^2 \bigr) \pmb{\biggr \rbrace} \biggr \rbrace \; .
  \end{split}
 \end{equation*}
%
\end{widetext}

\begin{widetext}
\section{Basic Integrals}\label{A-B}

For the calculation of the Feynman diagrams we need the following two-point integrals:
\begin{align}
J &= \int \frac{d^4 \ell}{(2\pi)^4} \frac{1}{\ell^2 (\ell+q)^2} = \frac{i}{32 \pi^2} \bigl \lbrack - 2 L \bigr \rbrack + \dots \label{B:1}\\
J_\mu &= \int \frac{d^4 \ell}{(2\pi)^4} \frac{\ell_\mu}{\ell^2 (\ell+q)^2} = \frac{i}{32 \pi^2} \bigl \lbrack q_\mu L \bigr \rbrack + \dots \label{B:2}\\
J_{\mu\nu} &= \int \frac{d^4 \ell}{(2\pi)^4} \frac{\ell_\mu \ell_\nu}{\ell^2 (\ell+q)^2} = \frac{i}{32 \pi^2} \biggl \lbrack q_\mu q_\nu \biggl( - \frac{2}{3} \ L \biggr) - q^2 \eta_{\mu\nu} \biggl ( - \frac{1}{6} \ L \biggr)   \biggr \rbrack + \dots \label{B:3} \\ 
J_{\mu\nu\alpha} &=  \int \frac{d^4 \ell}{(2\pi)^4} \frac{\ell_\mu \ell_\nu \ell_\alpha}{\ell^2 (\ell+q)^2} = \frac{i}{32 \pi^2} \biggl \lbrack q_\mu q_\nu q_\alpha \biggl ( \frac{1}{2} \, L \biggr)  - q^2 \bigl( q_\alpha \eta_{\mu\nu} + q_\mu \eta_{\nu\alpha} + q_\nu \eta_{\mu\alpha}\bigr) \biggl(  \frac{1}{12} \, L\biggr) \biggr \rbrack + \dots \label{B:4}\allowdisplaybreaks\\
J_{\mu\nu\alpha\beta} &=  \int \frac{d^4 \ell}{(2\pi)^4} \frac{\ell_\mu \ell_\nu \ell_\alpha \ell_\beta}{\ell^2 (\ell+q)^2} \notag \\
	& = \frac{i}{32 \pi^2} \biggl \lbrack q_\mu q_\nu q_\alpha q_\beta \biggl ( -\frac{2}{5} \, L \biggr)  - q^2 \bigl( q_\alpha q_\beta \eta_{\mu\nu} + q_\mu q_\beta \eta_{\nu\alpha} + q_\nu q_\alpha \eta_{\mu\beta} + q_\mu q_\beta \eta_{\nu\alpha} + q_\mu q_ \alpha \eta_{\nu\beta} \notag \\
	&\phantom{=\frac{i}{32 \pi^2}}  + q_\mu q_\nu \eta_{\alpha\beta}\bigr) \biggl(  -\frac{1}{20} \, L\biggr)  + q^4 \bigl( \eta_{\mu\nu}\eta_{\alpha\beta} + \eta_{\mu\alpha}\eta_{\nu \beta} + \eta_{\mu\beta}\eta_{\nu\alpha} \bigr) \biggl( - \frac{1}{120} \, L\biggr)\biggr \rbrack + \dots \, \label{B:5}
\end{align}
together with the three-point integrals
\begin{align}
I &= \int \frac{d^4 \ell}{(2\pi)^4} \frac{1}{\ell^2 (\ell +q)^2 \lbrack ( \ell + k)^2 - m^2 \rbrack} = \frac{i}{32 \pi^2 m^2} \bigl \lbrack - L - S \bigr \rbrack + \dots\label{B:6}\\
I_\mu &=  \int \frac{d^4 \ell}{(2\pi)^4} \frac{\ell_\mu}{\ell^2 (\ell +q)^2 \lbrack ( \ell + k)^2 - m^2 \rbrack} = \frac{i}{32 \pi^2 m^2}  \biggl \lbrace k_\mu \biggl \lbrack \biggl( -1 - \frac{1}{2}\frac{q^2}{m^2}\biggr) \, L - \frac{1}{4}\frac{q^2}{m^2} \, S \biggr \rbrack + q_\mu \biggl ( L + \frac{1}{2}\, S\biggr) \biggr \rbrace + \dots \label{B:7}\\
I_{\mu\nu} &= \int \frac{d^4 \ell}{(2\pi)^4} \frac{\ell_\mu \ell_\nu}{\ell^2 (\ell +q)^2 \lbrack ( \ell + k)^2 - m^2 \rbrack} \notag \\
	& = \frac{i}{32 \pi^2 m^2} \biggl \lbrace q_\mu q_\nu \biggl ( - L + \frac{3}{8} \, S \biggr) + k_\mu k_\nu \biggl( - \frac{1}{2} \frac{q^2}{m^2} \, L - \frac{1}{8}\frac{q^2}{m^2} \, S \biggr) + \bigl( q_\mu k_\nu + q_\nu k_\mu \bigr) \biggl \lbrack \biggl( \frac{1}{2} + \frac{1}{2}\frac{q^2}{m^2}\biggr) \, L + \frac{3}{16} \frac{q^2}{m^2} \, S\biggr \rbrack \notag \\
	&\phantom{=\frac{i}{32 \pi^2 m^2}} + q^2 \eta_{\mu\nu} \biggl( \frac{1}{4} \, L + \frac{1}{8} \, S\biggr) \biggr \rbrace + \dots \label{B:8}\allowdisplaybreaks\\
I_{\mu\nu\alpha} &= \int \frac{d^4 \ell}{(2\pi)^4} \frac{\ell_\mu \ell_\nu \ell_\alpha}{\ell^2 (\ell +q)^2 \lbrack ( \ell + k)^2 - m^2 \rbrack} \notag \\
	& = \frac{i}{32 \pi^2 m^2} \biggl \lbrace q_\mu q_\nu q_\alpha \biggl( L + \frac{5}{16} \, S \biggr) + k_\mu k_\nu k_\alpha \biggr( - \frac{1}{6}\frac{q^2}{m^2} \, L\biggr) + \bigl( q_\mu k_\nu k_\alpha + q_\nu k_\mu k_\alpha + q_\alpha k_\mu k_\nu \bigr) \biggl( \frac{1}{3} \frac{q^2}{m^2} \, L + \frac{1}{16} \frac{q^2}{m^2} \, S \biggr)\notag\allowdisplaybreaks \\
	&\phantom{= \frac{i}{32 \pi^2 m^2}} + \bigl( q_\mu q_\nu k_\alpha + q_\mu q_\alpha k_\nu + q_\nu q_\alpha k_\mu \bigr) \biggl \lbrack \biggl( - \frac{1}{3} - \frac{1}{2}\frac{q^2}{m^2} \biggr) \, L - \frac{5}{32} \frac{q^2}{m^2} \, S \biggr \rbrack \notag\allowdisplaybreaks \\
	&\phantom{= \frac{i}{32 \pi^2 m^2}} + \bigl( \eta_{\mu\nu}k_\alpha + \eta_{\mu\alpha}k_\nu + \eta_{\nu\alpha} k_\mu \bigr) \biggl( \frac{1}{12} \, q^2 L\biggr) + \bigl( \eta_{\mu\nu} q_\alpha + \eta_{\mu\alpha} q_\nu + \eta_{\nu\alpha} q_\mu \bigr) \biggl( - \frac{1}{6} \, q^2 L - \frac{1}{16} \, q^2 S \biggr) \biggr \rbrace + \dots \label{B:9}
\end{align}
where $L = \log (-q^2)$ and $S = \frac{\pi^2 m}{\sqrt{-q^2}}$, and the external momenta
satisfy $k^2=m^2$ and $k \cdot q = q^2 / 2$. 
Here, we have only quoted the leading nonanalytical terms. 
The ellipses denote higher order nonanalytical contributions and analytical terms. 
We have checked these integrals independently by explicit (manual) calculation and using {\sc Mathematica}.
%

\subsection{Reductions of the Four-Point Integrals}

For the box diagrams the following four-point integrals are needed:
\begin{align}
 K &= \int \frac{d^4 \ell}{(2 \pi)^4} \frac{1}{\ell^2 (\ell +q )^2 \lbrack (\ell+ k)^2 -m_1^2 \rbrack \lbrack (\ell - p)^2 -m_2^2 \rbrack} = \frac{i}{16 \pi^2 m_1 m_2 q^2} \biggl \lbrack \biggl( 1- \frac{w}{3 m_1 m_2} \biggr) \, L \biggr \rbrack + \dots\label{B:10}\\
 K' &= \int \frac{d^4 \ell}{(2 \pi)^4} \frac{1}{\ell^2 (\ell +q )^2 \lbrack (\ell+ k)^2 -m_1^2 \rbrack \lbrack (\ell + p')^2 -m_2^2 \rbrack} = \frac{i}{16 \pi^2 m_1 m_2 q^2} \biggl \lbrack \biggl( - 1+ \frac{W}{3 m_1 m_2} \biggr) \, L \biggr \rbrack + \dots\label{B:11} \; .
\end{align}
In these equations  $w = (k \cdot p) - m_1 m_2$ and $W = (k \cdot p') - m_1 m_2$,
which satisfy $W - w = k \cdot (p' - p) = (k \cdot q) = q^2 / 2$, \cite{Bje02, But06}. 

For on-shell scalar particles one has  the following identities:
\begin{equation}
\ell \cdot q =  \frac{1}{2}\bigl \lbrack (\ell+ q)^2 - q^2 - \ell^2 \bigr \rbrack \, , \quad  \ell \cdot k = \frac{1}{2} \bigl \lbrace \bigl \lbrack (\ell + k)^2 - m_1^2 \bigr \rbrack - \ell^2 \bigr \rbrace \; \text{and} \quad \ell \cdot p = - \frac{1}{2} \bigl \lbrace \bigl \lbrack (\ell - p )^2 - m_2^2 \bigr \rbrack - \ell^2 \bigr \rbrace 
\end{equation}
Since the terms with $(\ell + q)^2$ and $\ell^2$ do not contribute to 
 nonanalytical terms \cite{Don94a, Don94b, Bje02, Bje03, Akh97, Akh06, But06},
 we obtain for the four-point functions the simplifications/reductions
\begin{align}
 \int \frac{d^4 \ell}{(2 \pi)^4} &\frac{\ell \cdot q}{ \ell^2 (\ell +q)^2 \lbrack (\ell + k)^2 - m_1^2 \rbrack \lbrack ( \ell - p)^2 - m_2^2 \rbrack} = \frac{1}{2} \int \frac{d^4 \ell}{(2 \pi)^4} \frac{(\ell + q)^2 - q^2 - \ell^2}{ \ell^2 (\ell +q)^2 \lbrack (\ell + k)^2 - m_1^2 \rbrack \lbrack ( \ell - p)^2 - m_2^2 \rbrack}\notag \\
	&\longrightarrow \frac{-q^2}{2} \int \frac{d^4 \ell}{(2 \pi)^4} \frac{1 }{ \ell^2 (\ell +q)^2 \lbrack (\ell + k)^2 - m_1^2 \rbrack \lbrack ( \ell - p)^2 - m_2^2 \rbrack} = - \frac{q^2}{2} \ K \label{B:13} \; , \\
\int \frac{d^4 \ell}{(2 \pi)^4} &\frac{\ell \cdot k}{ \ell^2 (\ell +q)^2 \lbrack (\ell + k)^2 - m_1^2 \rbrack \lbrack ( \ell - p)^2 - m_2^2 \rbrack} = \frac{1}{2} \int \frac{d^4 \ell}{(2 \pi)^4} \frac{\lbrack (\ell + k)^2 - m_1^2\rbrack - \ell^2}{ \ell^2 (\ell +q)^2 \lbrack (\ell + k)^2 - m_1^2 \rbrack \lbrack ( \ell - p)^2 - m_2^2 \rbrack}\notag \\
&\longrightarrow \frac{1}{2} \int \frac{d^4 \ell}{(2 \pi)^4} \frac{1 }{ \ell^2 (\ell +q)^2 \lbrack ( \ell - p)^2 - m_2^2 \rbrack} = \frac{1}{2} \ I (m_2 , p) \label{B:14}\\
\int \frac{d^4 \ell}{(2 \pi)^4} &\frac{\ell \cdot p}{ \ell^2 (\ell +q)^2 \lbrack (\ell + k)^2 - m_1^2 \rbrack \lbrack ( \ell - p)^2 - m_2^2 \rbrack} = - \frac{1}{2} \int \frac{d^4 \ell}{(2 \pi)^4} \frac{\lbrack (\ell - p)^2 - m_2^2\rbrack - \ell^2}{ \ell^2 (\ell +q)^2 \lbrack (\ell + k)^2 - m_1^2 \rbrack \lbrack ( \ell - p)^2 - m_2^2 \rbrack}\notag \\
&\longrightarrow -\frac{1}{2} \int \frac{d^4 \ell}{(2 \pi)^4} \frac{1 }{ \ell^2 (\ell +q)^2 \lbrack ( \ell + k)^2 - m_1^2 \rbrack} =  - \frac{1}{2} \ I (m_1 , k) \label{B:15}
\end{align}
The four-point functions can thus be reduced to three- or two-point functions,
which simplifies the calculation significantly \cite{Bje02, Bje03, But06}. 

\end{widetext}

\twocolumngrid
%
\subsection{Constraints for the Nonanalytical Terms}

The following constraints for the nonanalytical terms are useful
\begin{equation}\label{B:16}
  J_{\mu\nu}\eta^{\mu\nu} = I_{\mu\nu} \eta^{\mu\nu} = I_{\mu\nu\alpha}\eta^{\mu\nu} = 0 \;  , 
\end{equation}
\begin{equation}\label{B:17}
\begin{split}
 J_\mu q^\mu &= - \frac{q^2}{2} \, J \; , \quad  J_{\mu\nu} q^\nu = - \frac{q^2}{2} \, J \; , \\
I_\mu q^\mu &= - \frac{q^2}{2} \, I \; , \quad  I_{\mu\nu} q^\nu = - \frac{q^2}{2} \, I_\mu \; ,  \\
I_{\mu\nu\alpha} q^\alpha &= - \frac{q^2}{2}\, I_{\mu\nu} \; ,\\
 \end{split}
\end{equation}
\begin{equation}\label{B:18}
 I_\mu k^\mu =\frac{1}{2} \, J \, , \; I_{\mu\nu} k^\nu = \frac{1}{2} \, k_\mu \, , \; I_{\mu\nu\alpha} k^\alpha = \frac{1}{2} \, k_{\mu\nu} \; .
\end{equation}
%
%

\section{Fourier Transformations}\label{A-C}

The basic Fourier integrals read
\begin{align}
 \int \frac{d^3 q}{(2 \pi)^3} \, e^{i \vec q \cdot \vec r} \ \frac{1}{|\ \vec q\ |^2} &= \frac{1}{4 \pi \ r}\label{C:1}\; , \\
\int \frac{d^3 q}{(2 \pi)^3} \, e^{i \vec q \cdot \vec r} \ \frac{1}{| \ \vec q \ |} &= \frac{1}{2 \pi^2 \ r^2}\label{C:2}\\
\intertext{and}
\int \frac{d^3 q}{(2 \pi)^3} \, e^{i \vec q \cdot \vec r} \ \log (\vec q^{\ 2}) &= - \frac{1}{2 \pi \ r^3} \; . \label{C:3}
\end{align}
%

\onecolumngrid

\section{Formal Expressions}\label{A-D}

Here we collect the formal expressions for the scattering amplitudes,
following from applying the Feynman rules given above to the various
diagrams. We only quote those results which
have not already been given in the body of the text:

\begin{itemize}
\item Figs~\ref{VerCorS1}(a) and (b):
\begin{align}
 i \M_{\ref{VerCorS1}(a)}(q) =& \tau_1^{\mu\nu}(k,k',m_1) \biggl \lbrack \frac{i \mathcal P_{\mu\nu\rho\sigma}}{q^2}\biggr \rbrack \int \frac{d^4 \ell}{(2 \pi)^4} \biggl \lbrack \frac{- i \ \eta_{\alpha\gamma}}{\ell^2}\biggr \rbrack  \tau_4^{\rho\sigma(\alpha\beta)}(\ell, \ell +q) \ \tau_2^\gamma (p,p-\ell)  \biggl \lbrack \frac{i}{(\ell - p )^2-m_2^2}\biggr \rbrack \notag\\
	&\cdot \tau_2^\delta (p-\ell, p',m_2) \biggl \lbrack \frac{- i \ \eta_{\delta\beta}}{(\ell +q)^2}\biggr \rbrack \; , \\
i \M_{\ref{VerCorS1}(c)}(q) =& \tau_1^{\mu\nu}(k,k',m_1) \biggl \lbrack \frac{i \mathcal P_{\mu\nu\rho\sigma}}{q^2}\biggr \rbrack \int \frac{d^4 \ell}{(2 \pi)^4} \tau_4^{\rho\sigma(\alpha\beta)} (\ell, \ell +q) \biggl \lbrack \frac{-i \ \eta_{\beta\delta}}{(\ell + q)^2}\biggr \rbrack \ \tau_5^{\delta\gamma}(p,p',m_2) \biggl \lbrack \frac{-i \ \eta_{\gamma\alpha}}{\ell^2}\biggr\rbrack \ , 
\end{align}
\item Figs~\ref{VerCorS2}(a) and (e):
\begin{align}
i \M_{\ref{VerCorS2}(a)}(q) =& \tau_2^\alpha (k,k',m_1) \biggl \lbrack \frac{- i \ \eta_{\alpha\beta}}{q^2}\biggr \rbrack   \int \frac{d^4 \ell}{(2\pi)^4}\ \tau_4^{\rho\sigma(\beta\gamma)}(q,\ell +q) \cdot \biggl \lbrack \frac{i \mathcal P_{\rho\sigma\mu\nu}}{\ell^2}\biggr \rbrack \tau_1^{\mu\nu}(p,p-\ell,m_2)\notag \\
	&\cdot \biggl \lbrack \frac{i}{(\ell - p)^2 -m_2^2} \biggr \rbrack \ \tau_2^\delta \ \biggl \lbrack \frac{- i \ \eta_{\delta\gamma}}{(\ell +q)^2}\biggr \rbrack \, , \\
i \M_{\ref{VerCorS2}(e)}(q) =& \tau_2^\alpha \biggl \lbrack \frac{-i \ \eta_{\alpha\beta}}{q^2}\biggr \rbrack \int \frac{d^4\ell}{(2\pi)^4} \tau_4^{\rho\sigma(\beta\gamma)}(q,\ell +q) \biggl \lbrack \frac{i \mathcal P_{\rho\sigma\mu\nu}}{\ell^2} \biggr \rbrack \tau^{\mu\nu(\delta)} (p,p',m_2) \biggl \lbrack \frac{-i \ \eta_{\delta\gamma}}{(\ell +q)^2}\biggr \rbrack \ ,
\end{align}
\item Figs~\ref{Tri} and \ref{TriS}:
\begin{align}
 i \M_{\ref{Tri}(a)}(q) =& \int \frac{d^4 \ell}{(2\pi)^4} \tau_1^{\mu\nu} \biggl \lbrack \frac{i \mathcal P_{\mu\nu\alpha\beta}}{\ell^2}\biggr \rbrack \tau_3^{\alpha\beta\gamma\delta}(k,k',m_2) \biggl \lbrack \frac{i \mathcal P_{\gamma\delta\rho\sigma}}{(\ell +q)^2} \biggr \rbrack \tau_1^{\rho\sigma}(k-\ell,k',m_1)\; ,\\
 i \M_{\ref{TriS}(a)}(q) =& \int \frac{d^4 \ell}{(2\pi)^4}\, \tau_2^\alpha (k, \ell +k,m_1) \biggl \lbrack \frac{-i \ \eta_{\alpha\beta}}{\ell^2}\biggr  \rbrack \tau_6^{\rho\sigma(\beta)}(p,p',m_2) \biggl \lbrack \frac{i \mathcal P_{\rho\sigma\mu\nu}}{(\ell +q)^2}\biggr \rbrack \tau_1^{\mu\nu}(\ell + k,k',m_1) \biggl \lbrack \frac{i}{(\ell + k)^2 -m_1^2}\biggr \rbrack\ , 
\end{align}
\item Figs~\ref{BCS}(a) and (c):
\begin{align}
 i \M_{\ref{BCS}(a)}(q) =& \int \frac{d^4 \ell}{(2 \pi)^4} \tau_2^{\alpha}(k,k + \ell , e_1) \biggl \lbrack \frac{- i \ \eta_{\alpha\beta}}{\ell^2}\biggr \rbrack  \tau_2^{\beta}(p, p - \ell,e_2) \biggl \lbrack \frac{i}{(\ell - p)^2 -m_2^2}\biggr \rbrack \tau_1^{\rho\sigma}(p-\ell, p' , m_2)\notag \\
	&\cdot  \biggl\lbrack \frac{i \mathcal P_{\rho\sigma\mu\nu}}{(\ell + q)^2}\biggr \rbrack  \tau_1^{\mu\nu}(\ell +k, k',m_1) \biggl \lbrack \frac{i}{(\ell +k)^2 -m_1^2}\biggr \rbrack \\
 i \M_{\ref{BCS}(c)} (q)&= \int \frac{d^4 \ell}{(2\pi)^4} \tau_2^\alpha (k, \ell +k, e_1) \biggl \lbrack \frac{-i \ \eta_{\alpha\beta}}{\ell^2} \biggr \rbrack \tau_2^\beta (p' + \ell, p', e_2 ) \biggl \lbrack \frac{i}{(\ell + p')^2 - m_2^2}\biggr \rbrack \tau_1^{\rho\sigma} (p, p' + \ell, m_2)\notag\\
	&\cdot  \biggl \lbrack \frac{i \mathcal P_{\rho\sigma\mu\nu}}{(\ell +q)^2} \biggr \rbrack  \tau_1^{\mu\nu} (\ell + k, k',m_1)\biggl \lbrack \frac{i}{(\ell + k)^2 -m_1^2}\biggr \rbrack \; ,
\end{align}
\item  Figs~\ref{CD}(a) and (b):
\begin{align}
 i M_{\ref{CD}(a)}(q) =& \frac{1}{2!}\int \frac{d^4 \ell}{(2 \pi)^4} \ \tau_3^{\mu\nu\alpha\beta} (k,k',m_1) \biggl \lbrack \frac{i \mathcal P_{\mu\nu\rho\sigma}}{\ell^2}\biggr \rbrack \tau_3^{\rho\sigma\gamma\delta} (p,p',m_2) \biggl \lbrack \frac{i \mathcal P_{\gamma\delta\alpha\beta}}{(\ell + q)^2}\biggr \rbrack \; ,\\
 i M_{\ref{CD}(b)}(q) =& \int \frac{d^4 \ell}{(2\pi)^4} \ \tau_6^{\mu\nu(\alpha)} (k,k',e_1) \biggl \lbrack \frac{- i \eta_{\alpha\beta}}{\ell^2}\biggr \rbrack \tau_6^{\rho\sigma(\beta)} (p,p',e_2) \biggl \lbrack \frac{i \mathcal P_{\rho\sigma\mu\nu}}{(\ell + q)^2}\biggr \rbrack \; .
\end{align}
\item Figs~\ref{VacP}(c), (d) and \ref{VacPS}:
\begin{align}
 i \M_{\ref{VacP}(c)} (q) =& \tau_1^{\mu\nu} (k,k',m_1) \biggl \lbrack \frac{i \mathcal P_{\mu\nu\alpha\beta}}{q^2} \biggr \rbrack  \bar\Pi^{\alpha\beta, \gamma \delta}\biggl \lbrack \frac{i \mathcal P_{\gamma\delta\rho\sigma}}{q^2}\biggr \rbrack \tau_1^{\rho\sigma} (p,p',m_2) \; ,\\
 i \M_{\ref{VacP}(d)}(q) =& \tau_1^{\mu\nu} (k,k',m_1) \biggl \lbrack \frac{i \mathcal P_{\mu\nu\alpha\beta}}{q^2}\biggr \rbrack \hat\Pi^{\alpha\beta, \gamma \delta} (q) \biggl \lbrack \frac{i \mathcal P_{\gamma\delta\rho\sigma}}{q}\biggr \rbrack \tau_1^{\rho\sigma} (p,p',m_2) \; ,\\
i \M_{\ref{VacPS}}(q)=& \tau_1^\delta (k,k',e_1) \biggl \lbrack \frac{-i \ \eta_{\delta\gamma}}{q^2}\biggr \rbrack \Pi^{\gamma,\tau} \biggl \lbrack \frac{-i \ \eta_{\tau\lambda}}{q^2}\biggr \rbrack \tau_1^\lambda (p,p',e_2)
\end{align}
\end{itemize}

\twocolumngrid
%

\end{document}